\documentstyle[aps,prb,multicol,epsf]{revtex} 
\begin{document} 
\widetext
\title{\bf  Superfluidity of bosons on a deformable lattice}

\author{G.~Jackeli\cite{byline} and J. ~Ranninger}

\address{Centre de
Recherches sur les Tr\`es Basses Temp\'eratures, 
Laboratoire
Associ\'e \'a l'Universit\'e Joseph Fourier, 
\\ Centre National de la
Recherche Scientifique, BP 166, 38042, Grenoble 
C\'edex 9, France}

\date{\today} 
\maketitle 
\draft 
\begin{abstract}

We study the superfluid properties of a system of interacting bosons 
on a lattice which, moreover, are  coupled to the  
vibrational modes of this lattice, treated here in terms of Einstein
phonon model. The ground state corresponds 
to two  correlated condensates: that of the bosons and that of the phonons. 
Two competing effects determine the common collective soundwave-like  
mode with sound velocity $v$, arising from gauge symmetry breaking:
i) The sound velocity $v_0$ (corresponding to a weakly interacting Bose 
system on a rigid lattice) in the lowest order approximation is reduced 
due to reduction of the repulsive boson-boson interaction, arising from 
the attractive part of phonon mediated interaction in the static limit. 
ii) the second order correction to the sound velocity is enhanced  as 
compared to the one of bosons on a rigid lattice when the the boson-phonon 
interaction is switched on due to the retarded nature of phonon mediated
interaction. The overall effect is that the sound velocity is practically
unaffected by the coupling with phonons, indicating the robustness of the
superfluid state.
The induction of a coherent state in the phonon system, driven by the
condensation of the bosons could be of experimental significance, permitting 
spectroscopic detections of  superfluid properties of the bosons.
Our results are based on an extension of the Beliaev - 
Popov formalism for a weakly interacting Bose gas on a rigid lattice to 
that  on a deformable lattice with which it interacts.
\end{abstract}


\begin{multicols}{2}

\narrowtext
\section{Introduction}

The weakly interacting Bose gas has been studied theoretically in 
great deal over the past fifty years;  mainly in view of understanding 
the superfluid properties of $^4\rm{He}$ and its rich phase diagram in the 
temperature - pressure parameter space. 
These studies provided a qualitative description of the low temperature 
collective soundwave like spectrum, the depletion of the condensate, 
etc.\cite{GRB} 
In more recent years attempts have been made to examine other systems than 
$^4\rm{He}$ which potentially could show such superfluid properties. The 
superfluidity of excitons in semiconducting materials such as 
$\rm{Cu}_2\rm{O}$, 
presents one of those novel systems.\cite{EXEXP} 
There, the interaction between 
the excitons with acoustic phonons seems to play a key role in 
establishing such a superfluidity of the bosonic excitons. In fact 
it involves the appearance of a coherent crystal 
displacement field which enables a moving exciton-phonon 
condensate.\cite{Loutsenko-97} 

In the present work we want to address ourselves to a general situation and 
consider the system in which the itinerant bosons, apart from an intrinsic 
repulsion between them, are coupled to some   bosonic degrees of 
freedom, which for brevity we shall term {\it phonons}. For instance, 
this situation
might be realized in short coherence length superconductors  
which are controlled by the fluctuations of the phase of the order 
parameter rather than of its amplitude. In such systems one expects 
relatively long lived local pairs of electrons which can be considered 
as  hard-core Bosons on a lattice and  which are coupled to the dynamical  
deformations of this underlying lattice. 
Ion-channelling experiments\cite{Ionexp} on High-$T_{c}$ cuprates
 (showing a drastic increase in the critical 
angle for Rutherford back-scattering just below $T_c$), as well as 
optical absorption measurements\cite{Raman} (showing a substantial 
increase in 
the phonon intensity just below $T_c$) indicate that, upon entering 
the superconducting state, the uncorrelated motion of the local lattice 
vibrations might get correlated, resulting in the emergence of an 
acoustic branch in the phonon spectrum. 

One also might think about systems, in which spin or pseudo-spin (orbital)
degrees are coupled to the lattice. If, upon lowering the temperature,
a long-range order due to the continuous symmetry breaking 
occurs in those systems, while the symmetry restoring variable is 
being coupled to the phonons, then the system will show  the same physics 
associated with macroscopic quantum effects as the model we are going to deal 
with.

The study presented in the present paper is, to our knowledge, 
the first generalization of the known field-theoretical treatment of 
the weakly interacting dilute Bose gas  when the effective
two-body potential is supplemented by a time dependent retarded 
part stemming from the phonon mediated retarded interaction. 
 
We shall study here the feasibility of such a phenomenon on the basis of a 
simple model of itinerant bosons on a deformable lattice.
In the present work we consider lattices susceptible of local lattice deformations. Such local deformations occur in the systems built up of molecular units and manifest themselves as dynamical deformations which are \`a priori spatially uncorrelated and are described by the Einstein phonon model.
Generalization of such a scenario to globally deformable lattices, described by acoustic phonon modes, will be briefly mentioned in the concluding part of the paper and will be subject of a future publication.  
In 
section 2 we shall define the model and study it within the standard 
Bogoliubov scheme\cite{Bogoliubov} which we  extend to two interacting 
boson fields: that of the itinerant bosons (the local electron pairs or 
electron-hole pairs respectively) and that of representing the 
lattice degrees of freedom   
(the optical  or acoustic phonons respectively). This 
approximation treats the scattering processes among condensate  
quasi-particles and between condensate and out-of-condensate quasi-particles, 
but neglects scattering among out-of-condensate  
quasi-particles. These latter processes are important if the system 
are not very dilute and if the temperature is finite. A consistent 
theory which can satisfactorily deal with that situation 
is the so called Beliaev-Popov theory\cite{B,P}, which, in section 3, 
we shall generalize from a weakly interacting Bose gas on a rigid lattice 
to one where also the interaction between the bosons and 
the lattice vibrations are taken into account. In section 4 we address 
the question of the renormalization of the sound velocity of the collective 
modes in the hydrodynamic regime at temperature zero. We shall illustrate 
the subtle compensation between two competing mechanisms arising from 
the boson-phonon coupling which while leading to an enhancement of the 
boson mass it also leads to an increased rigidity of the phase of 
the condensate. As a result the sound velocity is practically the same 
as that for a weakly interacting Bose gas on a rigid lattice. Concluding 
remarks and an outlook on further studies are presented in section 5.

\section{The model and its solution within a Bogoliubov scheme}

We consider a system of bosons on a lattice having a tight binding spectrum 
$\varepsilon_{\bf q}=zt[1-1/3(\cos q_{x}+\cos q_{y}+\cos q_{z})]$ 
(From now on we shall set $zt=1$, measuring all energies in units of the 
half bandwidth) which in the long wavelength limit is
given by $\varepsilon_{\bf q}= q^2/2M$ with  $M$ being the 
the bare boson  mass. 
The interaction between bosons is characterized by a coupling 
constant $g$ and that between the bosons and the phonons 
by $\alpha  \omega_0$, $\omega_0$ denoting the frequency of the optical 
phonon mode. The Hamiltonian for such a system is then given by
\begin{eqnarray}
H &=& \sum_{\bf q} \epsilon_{\bf q}b^{\dagger}_{\bf q} b_{\bf q} +  \omega_0 \sum_{\bf q}
(a^{\dagger}_{\bf q}a_{\bf q}+\frac{1}{2}) \nonumber \\
&+& \frac{g}{2N} \sum_{\bf k, k', p} b^{\dagger}_{\bf k} b^{\dagger}_{\bf k'} 
b_{\bf k'-p}b_{\bf k+p} \nonumber \\
&-& \frac{\alpha  \omega_0}{\sqrt N} \sum_{\bf k,q} 
b^{\dagger}_{\bf k} b_{\bf k+q}[a^{\dagger}_{\bf q} + a_{\bf -q}]
\label{H}
\end{eqnarray} 
 where $b^{({\dagger})}_{\bf k}$  and $a^{({\dagger})}_{\bf q}$ denote 
the boson and phonon annihilation (respectively creation) operators, 
$\epsilon_{\bf q}=\varepsilon_{\bf q}-\mu$ and $\mu$ being the chemical 
potential. 
Let us  now assume the existence of a condensed state not only for the 
bosons 
but also for the phonons and consequently make the Ansatz:
\begin{equation}
b_{\bf k} = \hat b_{\bf k} + \bar b \delta_{k,0}, \;  
a_{\bf k} = \hat a_{\bf k} + \bar a \delta_{k,0}
\label{shift}
\end{equation}
By substituting expressions (\ref{shift}) into the 
Hamiltonian (\ref{H}) and requiring that the terms linear 
in $\hat b_{\bf k}$ and $\hat a_{\bf q}$ vanish we find the 
following relations 
\begin{eqnarray}
\bar a &=& {\alpha \bar b^2 \over \sqrt N} = \alpha \sqrt N n_c \nonumber \\
n_c &=& {\bar b^2 \over N} \nonumber \\
\mu &\rightarrow& \bar{g}n_c,\;\;\bar{g}= g- 2 \alpha^2  \omega_0~.
\label{constraints}
\end{eqnarray}

finally yielding  the following Hamiltonian
\begin{equation}
H =  H_0 + H_{B-P} + H_{B-B}
\label{H_1}
\end{equation} 
and where 
\begin{eqnarray}
H_0 & = &\sum_{\bf q} {\bar \epsilon}_{\bf q} \hat b^{\dagger}_{\bf q} 
\hat b_{\bf q} + 
 \omega_0 \sum_{\bf q} \hat a^{\dagger}_{\bf q}\hat a_{\bf q} 
\nonumber \\
& + &  \frac{gn_c}{2} \sum_{\bf q}\left[ \hat b^{\dagger}_{\bf q} 
\hat b^{\dagger}_{\bf -q} + 
\hat b_{\bf q} \hat b_{\bf -q} \right] \nonumber \\
& - & \alpha\omega_0 \sqrt{n_c}  \sum_{\bf q} 
\left[ \hat b ^{\dagger}_{\bf q} + \hat b _{\bf -q} \right]
\left[ \hat a_{\bf q} + 
\hat a^{\dagger}_{-{\bf q}} \right]
\label{H_0}
\end{eqnarray}
with 
\begin{eqnarray}
{\bar \epsilon}_{\bf q} &=& \epsilon_{\bf q} + 2 g n_c - 2 \alpha^2  
\omega_0 n_c = 
\varepsilon_{\bf q} + gn_c.
\label{def}
\end{eqnarray}
The Hamiltonian $H_0$ describes the Bogoliubov quasi-particles which 
are hybridized with the phonons. In the condensed state the 
single particle excitations  hybridize  with the density fluctuations due 
to the broken gauge symmetry and give rise to the collective excitation
spectrum of the bosons with a sound-wave like spectrum. The phonons 
which initially couple to those density fluctuations hence get hybridized 
with those collective modes. The second term in the Hamiltonian 
Eq.~(\ref{H_1})  describes that coupling between the density of the 
out-of-condensate particles with the phonons.
\begin{equation}
H_{B-P} = -{\alpha  \omega_0 \over \sqrt N} \sum _{\bf k,q} 
\hat b^{\dagger}_{\bf k} 
\hat b_{\bf k+q}\left[ \hat a^{\dagger}_{\bf q} + 
\hat a_{\bf -q} \right].
\label{H_{B,P}}
\end{equation}
The last term in the Hamiltonian Eq.~(\ref{H_1}), $H_{B-B}$, describes 
the interaction between the out-of-condensate particles. 

Let us next define a  generalization of the standard Bogoliubov 
transformation
\begin{eqnarray}
\alpha_{\bf q} &=& u_{1,{\bf q}} \hat b_{\bf q} + 
u_{2,{\bf q}} \hat a_{\bf q} + v_{1,{\bf q}} \hat b^{\dagger}_{\bf -q} + 
v_{2,{\bf q}}\hat a^{\dagger}_{\bf -q} \nonumber \\
\beta_{\bf q} &=& \bar u_{1,{\bf q}} \hat b_{\bf q} + 
\bar u_{2,{\bf q}} \hat a_{\bf q} + \bar v_{1,{\bf q}} 
\hat b^{\dagger}_{\bf -q} + 
\bar v_{2,{\bf q}}\hat a^{\dagger}_{\bf -q} \nonumber \\
\alpha^{\dagger}_{\bf -q} &=& u_{1,{\bf q}} 
\hat b^{\dagger}_{\bf -q} + 
u_{2,{\bf q}} \hat a^{\dagger}_{\bf -q} + v_{1,{\bf q}} 
\hat b_{\bf q} + 
v_{2,{\bf q}}\hat a_{\bf q} \nonumber \\
\beta^{\dagger}_{\bf -q} &=& \bar u_{1,{\bf q}} 
\hat b^{\dagger}_{\bf -q} + 
\bar u_{2,{\bf q}} \hat a^{\dagger}_{\bf -q} + 
\bar v_{1,{\bf q}} \hat b_{\bf q} + 
\bar v_{2,{\bf q}}\hat a_{\bf q}.
\label{Bogol}
\end{eqnarray} 
With the requirement that the new operators satisfy the standard 
canonical commutation relations for Bosons, we obtain the following 
relations between the coefficients of this transformation
\begin{eqnarray} 
&[&\alpha_{\bf q},\alpha^{\dagger}_{\bf q}]=1\Rightarrow 
u^2_{1,{\bf q}}+u^2_{2,{\bf q}}-v^2_{1,{\bf q}}-v^2_{2,{\bf q}}  
\nonumber\\
&[&\beta_{\bf q},\beta^{\dagger}_{\bf q}]=1\Rightarrow 
\bar u^2_{1,{\bf q}}+\bar u^2_{2,{\bf q}}-\bar v^2_{1,{\bf q}}-
\bar v^2_{2,{\bf q}}
\label{cond}\\ 
&[&\alpha_{\bf q}, \beta^{\dagger}_{\bf q}]=0\Rightarrow
u_{1,{\bf q}} \bar u_{1,{\bf q}}+ u_{2,{\bf q}} \bar u_{2,{\bf q}}
- v_{1,{\bf q}} \bar v_{1,{\bf q}}-v_{2,{\bf q}} \bar v_{2,{\bf q}}  
\nonumber\\ 
&[&\alpha_{\bf q}, \beta_{\bf -q}]=0\Rightarrow
u_{1,{\bf q}} \bar v_{1,{\bf q}}+ u_{2,{\bf q}} \bar v_{2,{\bf q}}
- v_{1,{\bf q}} \bar u_{1,{\bf q}}-v_{2,{\bf q}} \bar u_{2,{\bf q}}.
\nonumber
\end{eqnarray}
The transformation, Eq.~(\ref{Bogol}), diagonalizes the Hamiltonian $H_0$, 
bringing it into the form 
\begin{equation}
\widetilde H_0 = \sum_{\bf q} 
\left( \omega_{1,{\bf q}} \alpha^{\dagger}_{\bf q}\alpha_{\bf q}  + 
 \omega_{2,{\bf q}} \beta^{\dagger}_{\bf q}\beta_{\bf q} \right).
\label{HD}
\end{equation}
The eigen-frequencies $\omega_{1,{\bf q}}$ and $\omega_{2,{\bf q}}$ 
as well as the coherence factors are determined in the usual way 
by diagonalizing the secular equation for the variables 
$\alpha_{\bf q}, \alpha^{\dagger}_{\bf -q}, \beta_{\bf q}, 
\beta^{\dagger}_{\bf -q}$ which leads to:
\begin{eqnarray}
\omega^2_{1,2,{\bf q}} &=& {E^2_{B {\bf q}} + \omega^2_0 \over 2} 
\nonumber \\
&\mp& \frac{1}{2} \sqrt {[E^2_{B {\bf q}} -  \omega^2_0]^2 + 
16 \alpha^2  \omega_0^3 \varepsilon_{{\bf q}}n_{c}}
\label{en} \\  
{u_1 \choose v_1} &=& {\omega_{1,{\bf q}} \pm \varepsilon_{{\bf q}} \over 
2 \sqrt{\omega_{1,{\bf q}}\varepsilon_{{\bf q}}}}
\sqrt{{ \omega^2_0 - \omega^2_{1,{\bf q}} \over 
\omega^2_{2,{\bf q}} -\omega^2_{1,{\bf q}}}}   \\
{u_2 \choose v_2} &=& { \omega_{1,{\bf q}}\pm \omega_0 \over 
2 \sqrt{\omega_{1,{\bf q}} \omega_0}} 
\sqrt{{E^2_{B {\bf q}}- \omega^2_{1,{\bf q}} \over 
\omega^2_{2,{\bf q}} -\omega^2_{1,{\bf q}}}}\\
{\bar u_1 \choose \bar v_1} &=& -{\omega_{2,{\bf q}} \pm 
\varepsilon_{0,{\bf q}} \over 
2 \sqrt{\omega_{2,{\bf q}}\varepsilon_{{\bf q}}}}
\sqrt{{ \omega^2_{2,{\bf q}}- \omega^2_0 \over 
\omega^2_{2,{\bf q}} -\omega^2_{1,{\bf q}}}}   \\
{\bar u_2 \choose \bar v_2} &=& { \omega_{2,{\bf q}}\pm \omega_0 \over 
2 \sqrt{\omega_{2,{\bf q}} \omega_0}} 
\sqrt{{\omega^2_{2,{\bf q}}-E^2_{B {\bf q}} \over 
\omega^2_{2,{\bf q}} -\omega^2_{1,{\bf q}}}}~,
\end{eqnarray}
and where $E_{B {\bf q}}=\sqrt{\varepsilon_{\bf q}[\varepsilon_{\bf q}+2gn_{c}]}$
denotes the spectrum of Bogoliubov quasi-particles in the absence of 
coupling with phonons.
 \begin{figure}
      \epsfysize=70mm
      \centerline{\epsffile{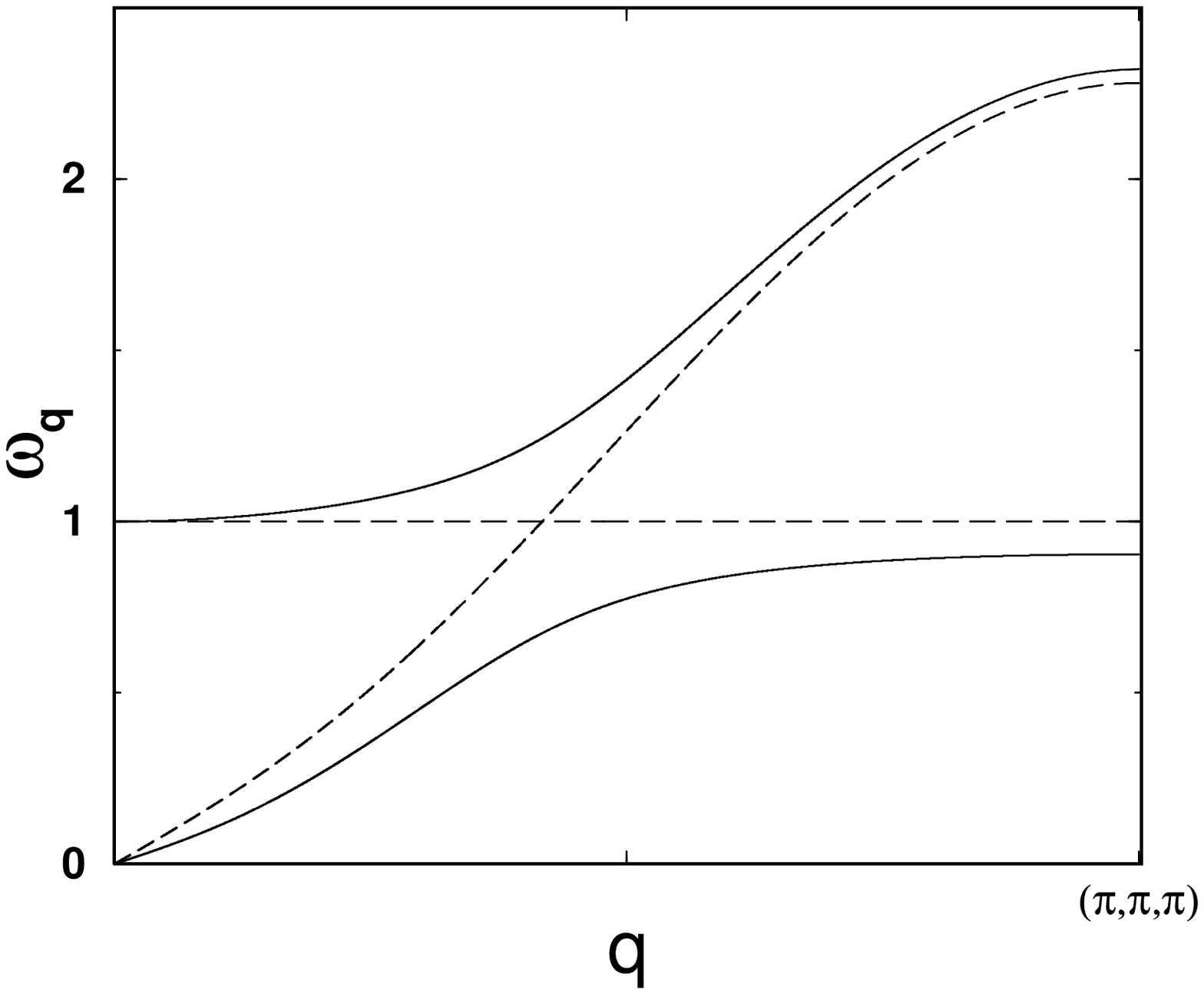}}
\caption{Sketch of the excitation spectra $\omega_{1,2,{\bf q}}$, 
Eq.~(\ref{en}), in the lowest order approximation (solid line).
The excitation spectra  in the absence of boson-phonon coupling is 
represented by dashed lines.}
\label{f1}
\end{figure}
In Fig. 1 we illustrate the two branches of eigen-frequencies 
$\omega_{1(2),{\bf q}}$. When $E_{B {\bf q}}$ is not close to the phonon 
frequency $\omega_{0}$, then 
the two normal modes consist of a Bogoliubov excitation
and a phonon mode; one of the modes being mainly a phonon and the other a 
Bogoliubov excitation. For $E_{B {\bf q}}$ close to $\omega_{0}$, 
neither mode is predominantly a phonon or a bosonic excitation. The mode 
which is a phonon (Bogoliubov excitation) for  
$E_{B {\bf q}} \ll \omega_{0}$ becomes a Bogoliubov excitation (phonon)
for $ \omega_{0}\ll E_{B {\bf q}}$. In the long-wavelength limit 
the lower branch  reduces to a sound-like dispersion
\begin{eqnarray}
\omega_{1,{\bf q}}\simeq v_{0}q, \;\; {\bf q}\to 0~,
\end{eqnarray}
with a characteristic sound velocity
\begin{eqnarray}
v_{0}=[(g-2\alpha^2\omega_{0})n_{c}/M]^{1/2}~.
\label{v0}
\end{eqnarray}
which is reduced in comparison to that where the boson-phonon 
coupling is absent. This reduction is determined by the static part 
of the attractive phonon mediated interaction.

The set of relations, Eqs. (11 - 15) represent the lowest order 
approximation to our problem, i.e.,
it takes into account collisions between condensate quasi-particles, 
between condensate and out-of-condensate quasi-particles, but 
totally neglects 
scattering among out-of-condensate quasi-particles as well as 
scattering out-of-condensate quasi-particles on phonons. 
As the concentration of 
quasi-particles increases these latter  become important. 
We therefore shall have to include in our study the effect of 
$H_{B-P}$ and $H_{B-B}$. In the next section, we shall generalize 
the Beliaev-Popov approximation (BPA)
\cite{B,P} for the case when the bosons, coupled together by the two-body 
potential, are moreover coupled by the phonon-mediated  retarded interaction.
Since in the BPA the second-order self-energies are built out of the
mean-field propagators we shall define all the Green's functions involved 
in our perturbation theory and give their expression within the mean-field 
approximation. These propagators can be defined  as the  elements 
of a matrix Green's function composed of a  four-component bosonic field
$\{ b_{\bf q},\; a_{\bf q},\;  b^{\dagger}_{\bf -q},\; 
a^{\dagger}_{\bf -q}\}$, the components of which are however 
related among each-other by symmetry relations. We define the 
canonical $2\times 2$ matrix 
Green's functions for bosons, with diagonal $(\cal G)$ and non-diagonal
$(\hat  {\cal G})$ components; the phonon Green's functions  $(\cal D)$ 
and the phonon-boson Green's function $(\cal H)$, that describes the 
hybridization of these two bosonic excitations in the condensed state 
\begin{eqnarray}
{\cal G}_{\bf q}(\omega)&=&\langle\langle
b_{\bf{q}}|b_{\bf{q}}^{\dagger}\rangle\rangle_{\omega}~,\;\;\;\;
\hat {\cal G}_{\bf q}(\omega)=\langle\langle
b_{\bf{q}}|b_{-\bf{q}}\rangle\rangle_{\omega}~,\nonumber\\
{\cal D}_{\bf q}(\omega)&=&\langle\langle
a_{\bf{q}}+a_{-\bf{q}}^{\dagger}|a_{\bf{q}}^{\dagger}+a_{-\bf{q}}
\rangle\rangle_{\omega}~,\\
\nonumber
\cal{H}_{\bf q}(\omega)&=&\langle\langle
b_{\bf{q}}|a_{\bf{q}}^{\dagger}+a_{-\bf{q}} \rangle\rangle_{\omega}~.
\label{GFdef}
\end{eqnarray}
The mean-field Green's functions, introduced above, 
are obtained from the diagonalized  Hamiltonian, Eq.~(\ref{HD}), with the 
help of the inverse canonical transformation 

\begin{eqnarray}
\hat b_{\bf q} &=& u_{1,{\bf q}} \alpha_{\bf q} + \bar u_{1,{\bf q}} 
\beta_{\bf q}
-  v_{1,{\bf q}} \alpha^{\dagger}_{\bf -q} - \bar v_{1,{\bf q}} 
\beta^{\dagger}_{\bf -q}
\nonumber \\
\hat a_{\bf q} &=& u_{2,{\bf q}} \alpha_{\bf q} + \bar u_{2,{\bf q}} 
\beta_{\bf q}
-  v_{2,{\bf q}} \alpha^{\dagger}_{\bf -q} - \bar v_{2,{\bf q}} 
\beta^{\dagger}_{\bf -q}
\nonumber \\
\hat b^{\dagger}_{\bf -q} &=& -v_{1,{\bf q}} \alpha_{\bf q} - 
\bar v_{1,{\bf q}} \beta_{\bf q}
+  u_{1,{\bf q}} \alpha^{\dagger}_{\bf -q} + \bar u_{1,{\bf q}} 
\beta^{\dagger}_{\bf -q}
\nonumber \\
\hat a^{\dagger}_{\bf -q} &=& -v_{2,{\bf q}} \alpha_{\bf q} - 
\bar v_{2,{\bf q}} \beta_{\bf q}
+  u_{2,{\bf q}} \alpha^{\dagger}_{\bf -q} + \bar u_{2,{\bf q}} 
\beta^{\dagger}_{\bf -q}.
\label{Bogi}
\end{eqnarray}

This leads to the following expressions for those mean-field 
Green's functions

\begin{eqnarray}
{\cal G}_{{\bf q},\omega}&=&\frac{u_{1,{\bf q}}^2}{\omega-
\omega_{1,{\bf q}}}-
\frac{v_{1,{\bf q}}^2}{\omega+\omega_{1,{\bf q}}}+
[\omega_{1,{\bf q}},u,v\rightarrow ,
\omega_{2,{\bf q}} {\bar u},{\bar v}]
~,\nonumber\\
\hat{\cal G}_{{\bf q},\omega}&=&-u_{1,{\bf q}}v_{1,{\bf q}}
\frac{2\omega_{1,{\bf q}}}{\omega^2-\omega_{1,{\bf q}}^2}+
[\omega_{1,{\bf q}},u,v\rightarrow ,
\omega_{2,{\bf q}}, {\bar u},{\bar v}]
~,\nonumber\\
{\cal D}_{{\bf q},\omega}&=&(u_{2,{\bf q}}-v_{2,{\bf q}})^2
\frac{2\omega_{1,{\bf q}}}{\omega^2-\omega_{1,{\bf q}}^2}
+
[\omega_{1,{\bf q}},u,v\rightarrow ,
\omega_{2,{\bf q}}, {\bar u},{\bar v}]
~,
\nonumber\\
\cal{H}_{{\bf q},\omega}&=&(u_{2,{\bf q}}-v_{2,{\bf q}})\left[
\frac{u_{1,{\bf q}}}{\omega-\omega_{1,{\bf q}}}+
\frac{v_{1,{\bf q}}}{\omega+\omega_{1,{\bf q}}}\right]\nonumber\\
&+&
[\omega_{1,{\bf q}},u,v\rightarrow ,
\omega_{2,{\bf q}}, {\bar u},{\bar v}]
~.
\label{GF}
\end{eqnarray}

In Fig. 2 we express graphically the phonon $\cal{D}$ and the boson-phonon
$\cal{H}$  Green's functions in terms of $\cal{G}$, the
normal and $\hat{\cal{G}}$, the anomalous component of the boson 
Green's functions. The choice of this graphical representation is 
convenient when 
constructing all the second-order self-energy diagrams, as will be
discussed in the next section. The full straight lines with two arrows 
in the same (opposite) direction represent the normal (anomalous) 
component of  boson Green's function.
The thin wavy and bold  wavy lines stand for the bare and the mean 
field Green's functions of the phonons. The boson-phonon Green's 
functions are  built up of a wavy line and an outgoing
straight line. The dot represents the boson-phonon vertex and the 
zig-zag line the condensate (given by a factor $n_{c}^{1/2}$). 
\begin{figure}
      \epsfysize=22mm
      \centerline{\epsffile{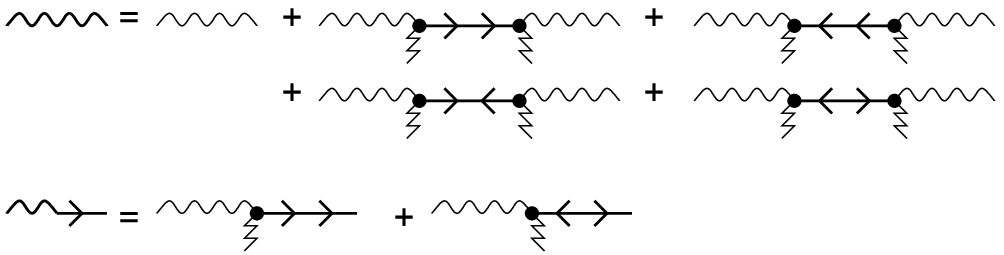}}
\caption{Graphical representation for the mean field Green's functions 
$\cal{D}$ and $\cal{H}$.}
\label{f2}
\end{figure}

\section{Generalized Beliaev-Popov Theory}

A consistent theory which goes beyond the mean-field approximation and 
takes into account collisions between out-of-condensate quasi-particles
was developed by Beliaev for a dilute Bose gas at $T=0$.\cite{B}
A generalization of this for finite temperature was done by Popov.\cite{P}
The Beliaev-Popov approximation is the next step beyond the lowest order
gapless theory and provides  a consistent second-order self-energy 
approximation for the  weakly interacting dilute Bose gas. 
It gives a gapless excitation spectrum with the  velocity equal
to the macroscopic speed of the sound at $T=0$,
in agreement with the findings by Gavoret and Nozi\`eres \cite{GN} which is 
valid to all orders in a perturbation theory treatment.
For details on the underling physics of the BPA we refer the reader 
to the review by Shi and Griffin.\cite{SG} 

\subsection{Effective interaction and higher-order contributions 
the self-energy}

In this section the BPA is generalized to the case where the quasi-particles,
interacting via a standard two-body potential, are moreover coupled 
by a phonon mediated retarded
interaction. To construct the second-order self-energy diagrams we define an 
effective interaction as the sum of a two-body $t$-matrix and a phonon 
mediated retarded interaction $\alpha^2\omega_{0}^2{\cal D}_{0}(\omega)$, 
with ${\cal D}_{0}(\omega)$ being given by the bare phonon Green's function.
\begin{figure}
      \epsfysize=30mm
      \centerline{\epsffile{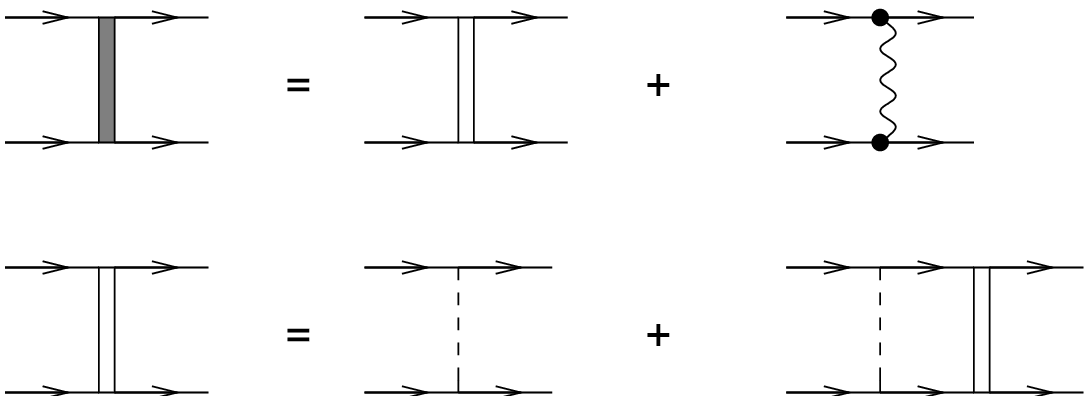}}
\caption{Graphical representation of the effective boson-boson interaction.}
\label{f3}
\end{figure}
The effective interaction is illustrated in Fig. \ref{f3}  as a hatched 
line. The dashed line stands for the bare interaction potential and the
non-hatched interaction line denotes the two-body $t$-matrix.  

In Fig. 4  we reproduce all the Beliaev-Popov diagrams for the normal 
$(\Sigma)$ and anomalous $(\hat\Sigma)$ components of self-energy.\cite{SG}
Each interaction line (hatched line) can be either a $t$-matrix
or a phonon propagator. The diagrams  are grouped in a way which is 
different from the conventional one. As it turns out  this type of 
grouping is more convenient when it comes to summing up in a compact 
form the subset of self-energy diagrams involving the phonon mediated 
interaction. We use the standard notation of Green's function lines having 
cuts which indicate diagrams which are already  included in the $t$-matrix 
and should be not counted again.
The mean field approximation, discussed in the previous section, retains only
the first two diagrams from the group $a1$ and the first diagram from the
group $a2$ in the normal and anomalous components of the self-energy
respectively. The $t$-matrix is then  being approximated by the $s$-wave 
scattering length.

Each second-order self-energy diagram consists of three contributions.
The first contribution arises when both effective interaction lines 
are replaced by 
the $t$-matrix. They have been extensively discussed in Ref.\onlinecite{SG}.
The only  difference between that and the  case we want to study here is 
the four pole structure
of the mean field Green's functions [see Eq.~(\ref{GF})]. The second 
contribution involves both the $t$-matrix and the phonon mediated interaction. 
Finally, the third contribution is exclusively due to the phonon 
mediated interaction. 
\end{multicols}
\widetext
\begin{figure}
      \epsfysize=150mm
      \centerline{\epsffile{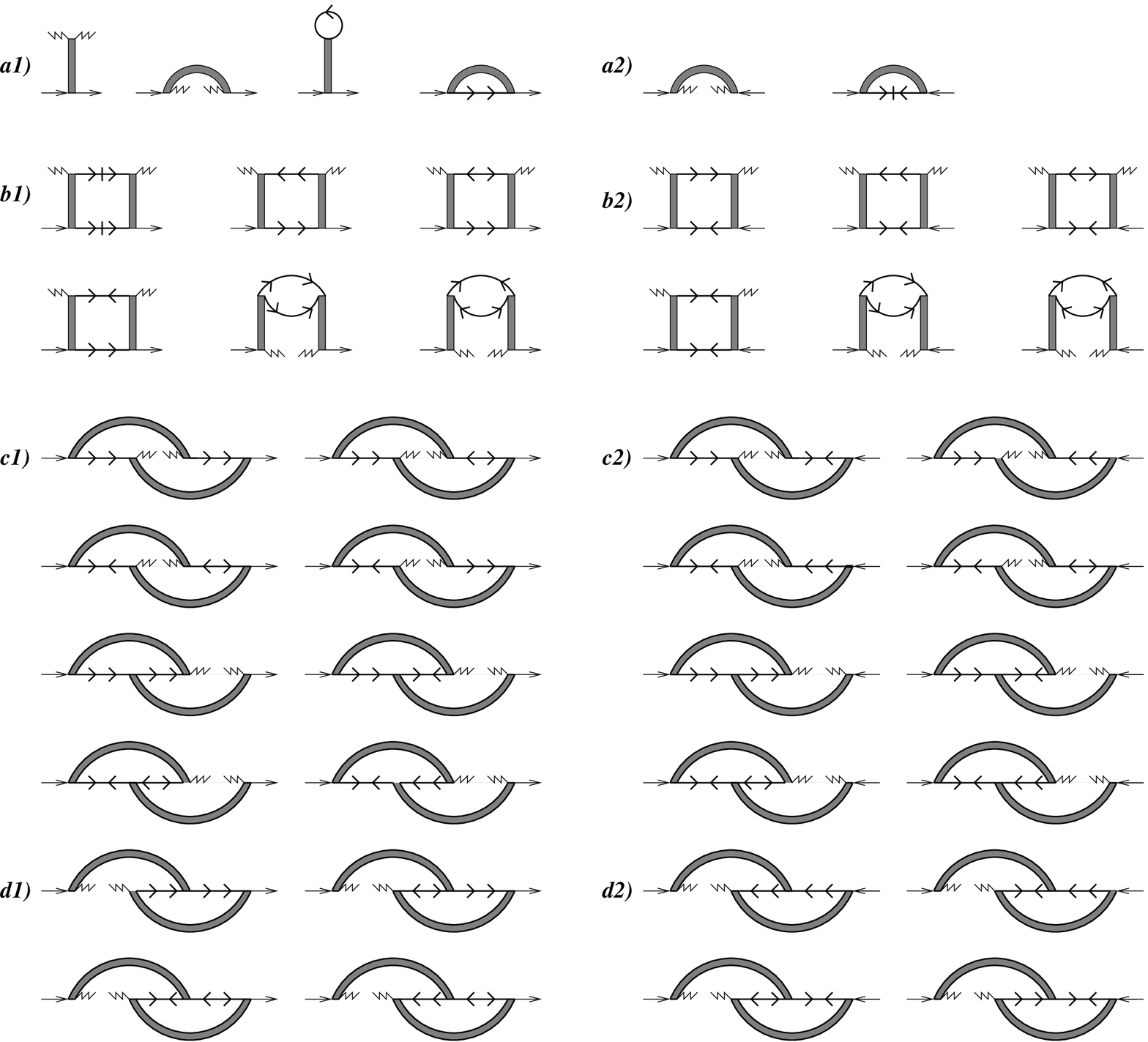}}
\caption{Graphical representation, up to second-order in the effective 
interaction of all contributions to the normal and anomalous self-energy. 
The hatched lines represents the effective interaction shown in Fig. 3.}
\label{f4}
\end{figure}
\begin{multicols}{2}

To sum up in the more compact form the subset of 
diagrams involving the boson-phonon coupling we use the graphical 
representation for the mean-field  Green's functions, derived in the
previous section [see Fig. 2].
We focus our discussion on the third subset of diagrams defined above.   
The second subset of diagrams, involving both types of interactions, 
are summed up using the same procedure. It turns out that the summation of 
the first four diagrams from group $b1$  together with the fourth diagram from
group $a1$ [see Fig. 4] is equivalent to  replacing the bare phonon propagator 
in the last diagram by the mean-field propagator. 
As a result one arrives 
at the first diagram shown in Fig. 5. 
Similarly, the summation of the first
four diagrams from the group $c1$ in Fig. 4 leads to the second diagram in
Fig. 5. The latter one represents the convolution of the Green's functions 
 describing the boson-phonon hybridization. The other diagrams in Fig. 5 
and the ones for the anomalous self-energy, represented in Fig. 6, are 
obtained in the similar way.
\begin{figure}
\epsfysize=30mm
\centerline{\epsffile{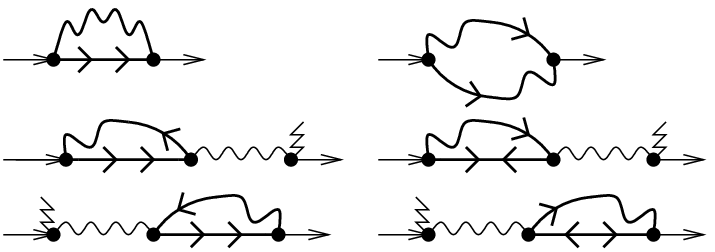}}
\caption{Graphical representation, up to second-order in the 
phonon mediated boson-boson 
interaction, of the normal self-energy $\Sigma$ due to the phonon 
mediated interaction.}
\label{f5}
\end{figure}
\begin{figure}
\epsfysize=30mm
\centerline{\epsffile{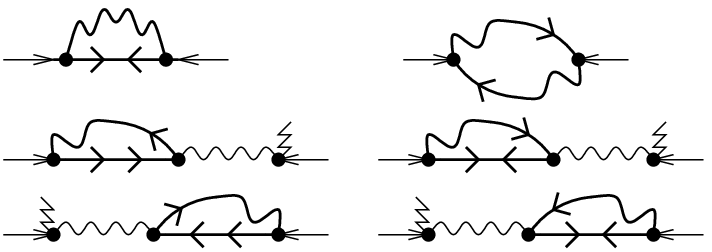}}
\caption{Graphical representation up to second-order in the 
phonon mediated boson-boson 
interaction, of the anomalous self-energy $\hat\Sigma$ due to the
phonon-mediated interaction.}
\label{f6}
\end{figure}

We would like to point out that, in the  present treatment,
the renormalization of the phonon mode as well as of all relevant vertex 
corrections is consistently taken into account up to 
the second-order in the effective interaction. 
The phonon mode is indeed renormalized
by  the density fluctuation of bosons.
In the condensed state the density response function consists of 
two contributions: a quasi-particle part and 
a part describing density fluctuation of out-of-condensate particles. 
The phonon renormalization due to the first contribution is
already present in the extended Bogoliubov scheme, leading to the 
renormalized (mean-field ) propagator of phonons.
By replacing the  effective interaction by phonon propagators, one verifies
that  the first diagram from group $a1$ in Fig. 4 
together with last two diagrams 
from group $b1$ describe the phonon mode  
renormalization arising from the density fluctuations
 of out-of-condensate particles. They are described
given by the ``particle-hole'' bubble diagrams built up from normal 
and anomalous mean-field (Bogoliubov) Green's functions of the bosons.
One farther verifies, that the diagrams from group $d1$ 
together with diagrams of identical topology from group $c1$ from Fig. 4
generates the vertex corrections to the boson-phonon as well as to the boson-boson vertex. For example, when all the effective interaction lines in these
diagrams  are replaced by the phonon mediated interaction,
one obtains the vertex corrections to the phonon-boson interaction.
Similarly appear  the corrections to the boson-phonon vertex 
due to the boson-boson interaction 
(these diagrams are generated by replacing one  effective interaction line
 by the boson-boson interaction line). 
When all the phonon lines in the above mentioned class of diagrams
 are replaced by  
boson-boson interaction lines one obtains the renormalization of the boson-boson interaction. 
In a similarly way we can generate 
  the diagrams  renormalizing the boson-boson interaction by the phonon-boson 
interaction.

The normal and anomalous components of the Boson 
Green's functions  are expressed in 
terms of the self-energy  via the Dyson-Beliaev equation
\begin{eqnarray}
{\cal G}_{{\bf q},\omega}&=&\frac{\omega+\epsilon_{\bf
q}+\Sigma_{{\bf q},-\omega}}{{\text D}_{{\bf q},{\omega}}}~,\;\;\;
\hat{ \cal G}_{{\bf q},\omega}=\frac{-\hat\Sigma_{{\bf
q},\omega}}{{\text D}_{{\bf q},{\omega}}}~,
\label{DB}
\end{eqnarray}
where $\epsilon_{\bf q}$ is the bare  excitation spectrum measured from 
the chemical potential $\mu$. In Eq.~( \ref{DB})  the following notations
are  introduced: 
\begin{eqnarray}
{\text D}_{{\bf q},{\omega}}&=&[\omega-{\cal A}_{{\bf q},\omega}]^2-
[\epsilon_{\bf q}+S_{{\bf
q},\omega}][\epsilon_{\bf q}+S_{{\bf q},\omega}+2\hat\Sigma_{{\bf
q},\omega}]~,\nonumber\\
{\cal A}_{{\bf q},\omega}&=&\frac{\Sigma_{{\bf q},\omega}-\Sigma_{{\bf
q},-\omega}}{2}~,\\
\label{DBnot}
{\cal S}_{{\bf q},\omega}&=&\frac{\Sigma_{{\bf q},\omega}+
\Sigma_{{\bf q},-\omega}-2\hat\Sigma_{{\bf
q},\omega}}{2}~.\nonumber
\end{eqnarray}
Here ${\cal A}_{{\bf q},\omega}$ is an antisymmetric function in $\omega$, 
while  ${\cal S}_{{\bf q},\omega}$ and  $\hat\Sigma_{{\bf q},\omega}$ are 
symmetric functions of $\omega$.  Invoking the Hugenholtz-Pines theorem 
(see the following sub-section) we have the relation 
$\mu=\Sigma_{0,0}-{\hat\Sigma}_{0,0}=S_{0.0}$.

The existence of the two branches in the  excitation spectrum, Eq.~(\ref{en}) 
leads to three different contributions to the self-energy, describing the
intra-band and inter-band scattering processes.  Based on the mean-field 
Green's functions, Eq.~(\ref{GF}),
we evaluate the above introduced dynamical quantities
at zero temperature and  arrive at the following result:
\begin{eqnarray}
{\cal A}_{{\bf q},\omega}&=&\frac{1}{N}\sum_{{\bf k},\alpha,\beta}
\frac{A_{\alpha,\beta}({\bf k},{\bf q})\omega}{\omega^2-(\omega_{\alpha,{\bf k}}+
\omega_{\beta,{\bf k}+{\bf q}})^2}~,\nonumber\\
{\cal S}_{{\bf q},\omega}&=&\frac{1}{N}\sum_{{\bf k},\alpha,\beta}
\frac{S_{\alpha,\beta}({\bf k},{\bf q})(\omega_{\alpha,{\bf k}}+
\omega_{\beta,{\bf k}+{\bf q}})}{\omega^2-(\omega_{\alpha,{\bf k}}+
\omega_{\beta,{\bf k}+{\bf q}})^2}~,\nonumber\\
{\hat\Sigma}_{{\bf q},\omega}&=&\frac{1}{N}\sum_{{\bf k},\alpha,\beta}
\frac{M_{\alpha,\beta}({\bf k},{\bf q})(\omega_{\alpha,{\bf k}}+
\omega_{\beta,{\bf k}+{\bf q}})}{\omega^2-(\omega_{\alpha,{\bf k}}+
\omega_{\beta,{\bf k}+{\bf q}})^2}~.
\label{SE}
\end{eqnarray}
$\alpha~(\beta)$ denotes the branch index, $\beta\leq\alpha=1,2$,
the corresponding vertices $A_{\alpha,\beta}$, $S_{\alpha,\beta}$,
and  $M_{\alpha,\beta}$ which are  expressed in terms of the coupling 
constants and coherence factors and are given in Appendix A.

The renormalized quasi-particle spectrum is given by the poles 
of the dressed boson Green's function,  or equivalently by the zeros
of ${\text D}_{{\bf q},{\omega}}$. In the next section we discuss
the long-wave length excitation spectrum of the system, focusing  on
the effect of boson-phonon coupling  on the sound-velocity renormalization.  

\subsection{The Hugenholtz-Pines theorem}

One of the fundamental results in the theory of Bose systems is the
Hugenholtz-Pines theorem\cite{HP} which relates the chemical potential to 
the self-energy in the condensed state via the relation
\begin{eqnarray}
\mu=\Sigma_{0,0}-\hat{\Sigma}_{0,0}~.
\label{mu}
\end{eqnarray}
This  theorem implies that the excitation
spectrum of the system is gapless. However in certain approximations,
for example the Hartree-Fock-Bogoliubov approximation (HFB), 
higher-order terms in the chemical potential are retained, while being 
neglecting in the self-energies. Such   approximations are not "consistent" 
to any given order in the coupling constants and
can  generate an unphysical gap in the excitation spectrum of the system.
Let us now  verify this point for our treatment.
We first  calculate the chemical potential from the Heisenberg
equations of motion for the Bose field (analogous to the  Gross-Pitaevskii 
equation)
and compare the result to  that obtained from the Hugenholtz-Pines theorem.
Our further treatment closely follows that of Ref.\onlinecite{GRIF}.

We start from the exact Heisenberg equation of motion for the Bose field
$b_{\bf q}$
\begin{eqnarray}
i{\dot b}_{\bf q}&=&[\varepsilon_{\bf q}-\mu]b_{\bf q}
+\frac{g}{N}\sum_{{\bf k},{\bf p}}
b^{\dagger}_{\bf k+p}b_{\bf k+q}b_{\bf p}\nonumber\\
&-&\frac{\alpha  \omega_0}{\sqrt N} \sum_{\bf k,q} 
b_{\bf k+q}[a^{\dagger}_{\bf k} + a_{\bf -k}].
\label{eom}
\end{eqnarray}
The next step is to separate out the condensate part as in Eq.~(\ref{shift})
and treat the boson interaction term in a self-consistent mean-field 
approximation (HFB approximation).\cite{GRIF} As a result we arrive at  
the following equation
\begin{eqnarray}
\mu{\bar b}&=&gn_{c}{\bar b}+\frac{g}{N}\sum_{\bf q}
\left[2\langle{\hat b}^{\dagger}_{\bf q}{\hat b}_{\bf q}\rangle
+ \langle{\hat b}^{\dagger}_{\bf q}{\hat b}_{\bf q}\rangle\right]
{\bar b}\nonumber\\
&-&2\alpha^2\omega_{0}n{\bar b}-\frac{\alpha\omega_{0}}{\sqrt{N}}
\sum_{\bf q}\langle {\hat b}_{\bf q}[{\hat a}^{\dagger}_{\bf q} + 
{\hat a}_{\bf -q}]\rangle~,
\label{cwf}
\end{eqnarray}
where ${\bar b}^2/N=n_{c}$ is the condensate fraction and $n$ is the boson 
density.
The above equation yields the following result for the chemical potential:
\begin{eqnarray}
\mu=\mu_{1}+\mu_{2},\;\mu_{1}=\bar{g}n,\;
\mu_{2}=g[\tilde{n}+\tilde{m}]
-\alpha\omega_{0}\tilde{h}~,
\label{mugp}
\end{eqnarray}
$\tilde{n}$,  $\tilde{m}$ denote the density and anomalous density of
out-of-condensate particles, respectively and  $\tilde{h}$ is
the strength of boson-phonon hybridization. These quantities are given by
\begin{eqnarray}
\tilde{n}&=&\frac{1}{N}\sum_{\bf q}
\langle{\hat b}^{\dagger}_{\bf q}{\hat b}_{\bf q}\rangle~,\;\;\;
\tilde{m}=\frac{1}{N}\sum_{\bf q}\langle{\hat b}_{\bf q}{\hat b}_{\bf q}
\rangle~, \nonumber\\
\tilde{h}&=&\frac{1}{\sqrt{n_c}N} \sum_{\bf q} 
\langle {\hat b}_{\bf q}[{\hat a}^{\dagger}_{\bf q} + 
{\hat a}_{\bf -q}]\rangle~.
\label{nmh}
\end{eqnarray}

So far we did not make any approximation for the 
boson-phonon interaction. We next  show that if all the above defined
averages are calculated self-consistently from the corresponding mean field
Green's functions, then the derived chemical potential Eq.~(\ref{mugp})
 will coincide with that obtained from the Hugenholtz-Pines theorem.

In Fig. 7 we represent graphically the  contributions
to the chemical potential, Eq.~(\ref{mugp}),  arising from  the coupling 
of the bosons with the phonons.
Putting this contribution together with the first contribution to $\mu_{2}$
in Eq.~(\ref{mugp}) we obtain $\mu_2=A+\hat{\large{A}}$.
$A$ and $\hat{A}$ are the contributions corresponding to the fourth diagram 
from the group $a1$ and the second diagram from the group $a2$, presented 
in Fig. 4.
\begin{figure}
      \epsfysize=12mm
      \centerline{\epsffile{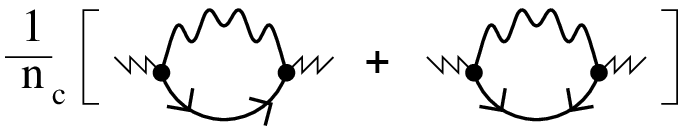}}
\caption{Graphical representation of second-order contributions to 
the chemical potential due to the coupling with phonons.}
\label{f7}
\end{figure} 
Now we calculate the chemical potential from the Hugenholtz-Pines theorem.
Examining the various diagrams of $\Sigma_{0,0}$ and  $\hat{\Sigma}_{0,0}$
we find that  most of them cancel pairwises, giving the following
result
\begin{eqnarray}
\mu_{\text{HP}}&=&\bar{g}n+g[\tilde{n}+\tilde{m}_{\text{R}}]+A-\hat{A}
+B~,\nonumber\\
\tilde{m}_{\text{R}}&=&\tilde{m}-\frac{gn_{c}}{N}\sum_{\bf q}\frac{1}
{2\varepsilon_{\bf q}}~.
\label{muhp}
\end{eqnarray}
$\tilde{m}_{\text{R}}$ denotes the "renormalized" anomalous density
\cite{GRIF} and
\begin{eqnarray*}
B=\frac{1}{N}\sum_{\bf q}\int d\omega n_{c}[g-\alpha^2\omega_{0}^2{\cal
D}_{0}(\omega)]^2\{{\cal G}_{\omega}{\cal G}_{-\omega}
-[\hat{\cal G}_{\omega}]^2\}~.
\end{eqnarray*}
Using the Dyson-Beliaev equation (\ref{DB}) we verify the following 
identities
\begin{eqnarray}
2\hat{\cal G}_{\omega}&=&n_{c}[g-\alpha^2\omega_{0}^2{\cal
D}_{0}(\omega)]\{{\cal G}_{\omega}{\cal G}_{-\omega}
-[\hat{\cal G}_{\omega}]^2\}~,\nonumber\\
B&=&\frac{2}{N}\sum_{\bf q}\int d\omega [g-\alpha^2\omega_{0}^2{\cal
D}_{0}(\omega)]
\hat{\cal G}_{\omega}=2\hat{A}~.
\label{B2}
\end{eqnarray}
Substituting the expressions, Eq.~(\ref{B2}) into Eq.~(\ref{muhp}) 
we see that
the chemical potential derived from the Hugenholtz-Pines theorem 
coincides with the
one obtained from the Heisenberg equations of motion, Eq.~(\ref{mugp}). 

\section{Results and discussion}

In the preceding section we have given the expressions for the self-energies
in the Beliaev-Popov approximation. We now use these results to
calculate the second-order corrections to the sound velocity and the chemical
potential.
 
\subsection{Depletion of the ground state}

We first discuss, the depletion of the ground state. From the Green's 
function Eq.~(\ref{GF}) we  determine the density of out-of-condensate 
particles $\tilde{n}$, and thus obtain the self-consistent equation 
for the density of particles in the condensate:
\begin{eqnarray}
n&=&n_{c}+\tilde{n},\; \tilde{n}=\frac{1}{N}\sum_{\bf k}[v_{1,{\bf k}}^2+{\bar v}_{1,{\bf
k}}^2]
\label{depl}\\
&=&\frac{1}{N}\sum_{\bf
k}\left\{\frac{[\bar{g}n_{c}+\varepsilon_{\bf k}][\omega_{0}+
\bar{E}_{{\bf k}}]}{2\bar{E}_{{\bf k}}
(\omega_{1,{\bf k}}+\omega_{2,{\bf k}})}+\frac{\alpha^2\omega_{0}n_{c}}
{\omega_{1,{\bf k}}+\omega_{2,{\bf k}}}-\frac{1}{2}\right\},\nonumber
\end{eqnarray}   
$\bar{E}_{ {\bf q}}=\sqrt{\varepsilon_{\bf q}[\varepsilon_{\bf
q}+2\bar{g}n_{c}]}$ is  the spectrum of Bogoliubov quasi-particles with
reduced scattering length $\bar{g}$ [see Eq.~(\ref{constraints})]. 
The self-consistent equation for 
the condensate fraction, Eq.~(\ref{depl}), is solved numerically. In Fig. 8 
the relative depletion $r=\tilde{n}/n$ as a function of density is
presented for various values of the boson-phonon coupling constant $\alpha$. 
As clearly evident from this figure, for vanishing $\alpha$, the relative 
depletion follows the canonical $\sqrt{n}$ behavior, which is  
characteristic for the dilute weakly interacting Bose gas. 
For a non-zero  coupling constant however, $r$ remains finite in the limit
$n\rightarrow 0$, indicating a linear in density contribution to the
depletion of the ground state.
\begin{figure}
\epsfysize=70mm
\centerline{\epsffile{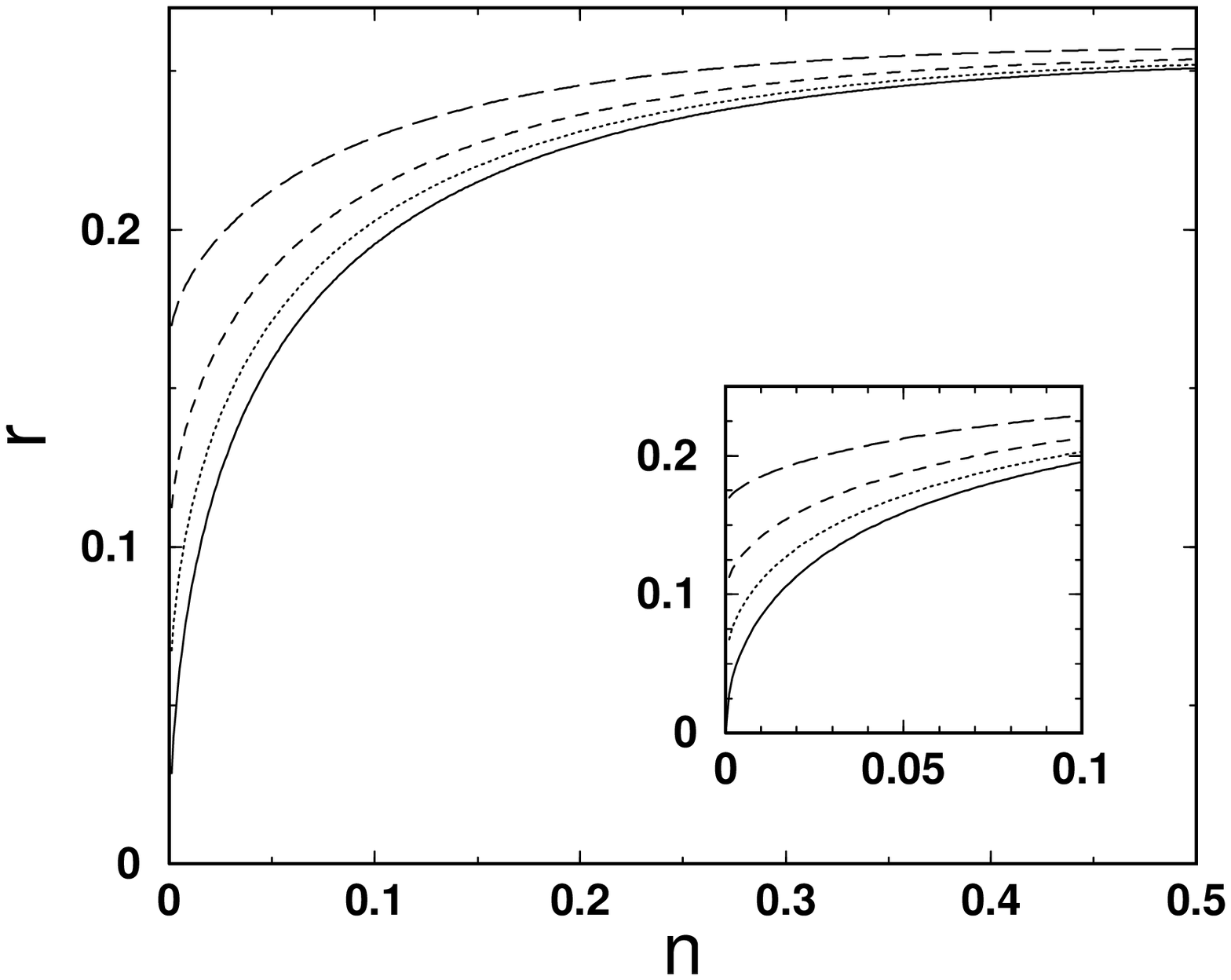}}
\caption{Relative depletion of the ground state $r=\tilde{n}/n$ as a 
function of density for different values of  coupling constants 
($\alpha=$0, 1, 1.5, and 2 --  solid line, dotted, dashed and long 
dashed lines, respectively) for $\omega_{0}=0.2$ and 
g=3.}
\label{f8}
\end{figure}
For $n\ll 1$, expanding the denominator in the last sum of 
Eq.~(\ref{depl}) and keeping the leading order  contributions in $n_{c}$
we arrive at the following expression  
\begin{eqnarray}
\tilde{n}\simeq\frac{1}{N}\sum_{\bf k}
\left[\frac{\bar{g}n_{c}+\varepsilon_{\bf k}-\bar{E}_{\bf k}}
{2\bar{E}_{\bf k}}
+\frac{\alpha^2\omega_{0}^2n_{c}}
{(\omega_{0}+\bar{E}_{\bf k})^2}\right]~.
\label{depl1}
\end{eqnarray}   
The first term in Eq.~(\ref{depl1}) gives rise to the canonical
$n_{c}^{3/2}$ contribution to the ground state depletion while the second
term  leads to the linear in density contribution, and is due 
to the boson-phonon coupling.

\subsection{The chemical potential}

The second order contribution to the chemical potential is given as
a sum of two contributions: $\mu_{2}=\mu_{b}+\mu_{p}$. The first one 
is due to the boson-boson interaction and the second one stems from 
the boson-phonon interaction:
\begin{eqnarray}
\mu_{b}&=&g(\tilde{n}+\tilde{m_{\text{R}}})=\frac{g}{2N}\sum_{\bf k}
\left[\frac{\varepsilon_{\bf k}}{\bar{E}_{\bf k}}\frac{\omega_0+
\bar{E}_{\bf k}}{\omega_{1,{\bf k}}+\omega_{2,{\bf
k}}}+\frac{gn_{c}}{\varepsilon_{\bf k}}-1\right],\nonumber\\
\mu_{p}&=&-\alpha\omega_{0}\tilde{h}=-\alpha^2\omega_{0}^2\sum_{\bf k}
\frac{\varepsilon_{\bf k}}{
\bar{E}_{\bf k}(\omega_{1,{\bf k}}+\omega_{2,{\bf
k}})}.
\label{mubp}
\end{eqnarray}
In the normal state, where $n_{c}=0$, the second contribution $\mu_{p}$ 
coincides with the density independent negative shift of the chemical 
potential due to the coupling with phonons. However, in the condensed 
state this contribution becomes density dependent, via the density 
dependence of the spectrum
in the condensed state. The density dependent part of this contribution
increases with increasing density and hence gives rise to an increase 
of the system's compressibility. In Fig. 9 we plot the density dependent 
part of this contribution ($\mu_{p}=\mu_{p}(n)-\mu_{0}(0)$)  for several 
coupling constants.
\begin{figure}
\epsfysize=70mm
\centerline{\epsffile{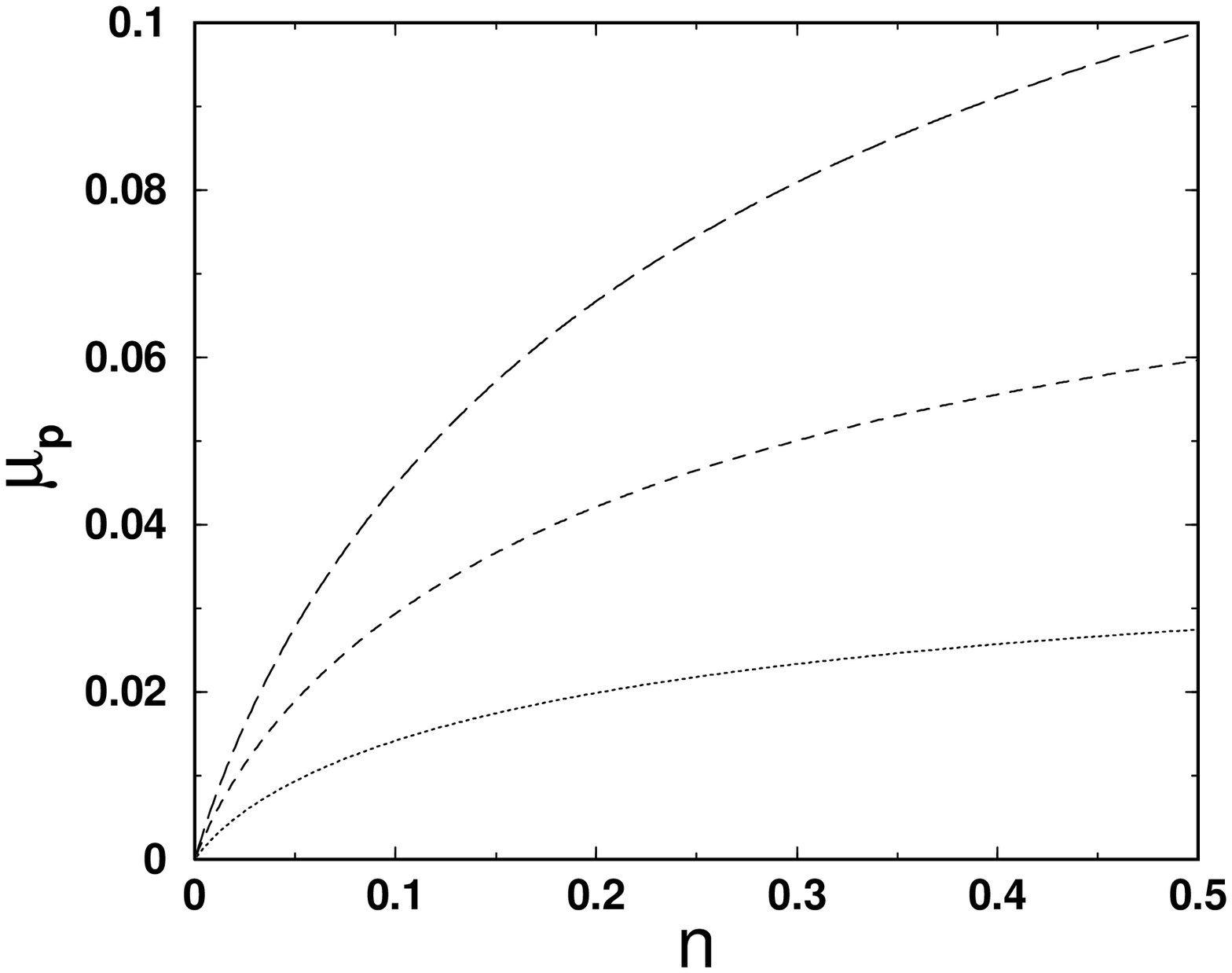}}
\caption{Second order contribution to the chemical potential due to the
boson-phonon coupling ($\mu_{p}$) versus density (n).
($\alpha=$ 1, 1.5, and 2 -- dotted, dashed and long dashed lines,
respectively;  $\omega_{0}=0.2$ and g=3.}
\label{f9}
\end{figure}
The total second order correction to the chemical potential is presented in
Fig. 10 which also shows its increase with $\alpha$. We would like to emphasize
that for $\alpha=0$ and for $n\rightarrow 0$ the second order correction to
the chemical potential shows the canonical $n^{3/2}$ behavior, while for
large $\alpha$ the slope of $\mu_{2}$ is almost linear [see inset on Fig. 10].
\begin{figure}
      \epsfysize=70mm
      \centerline{\epsffile{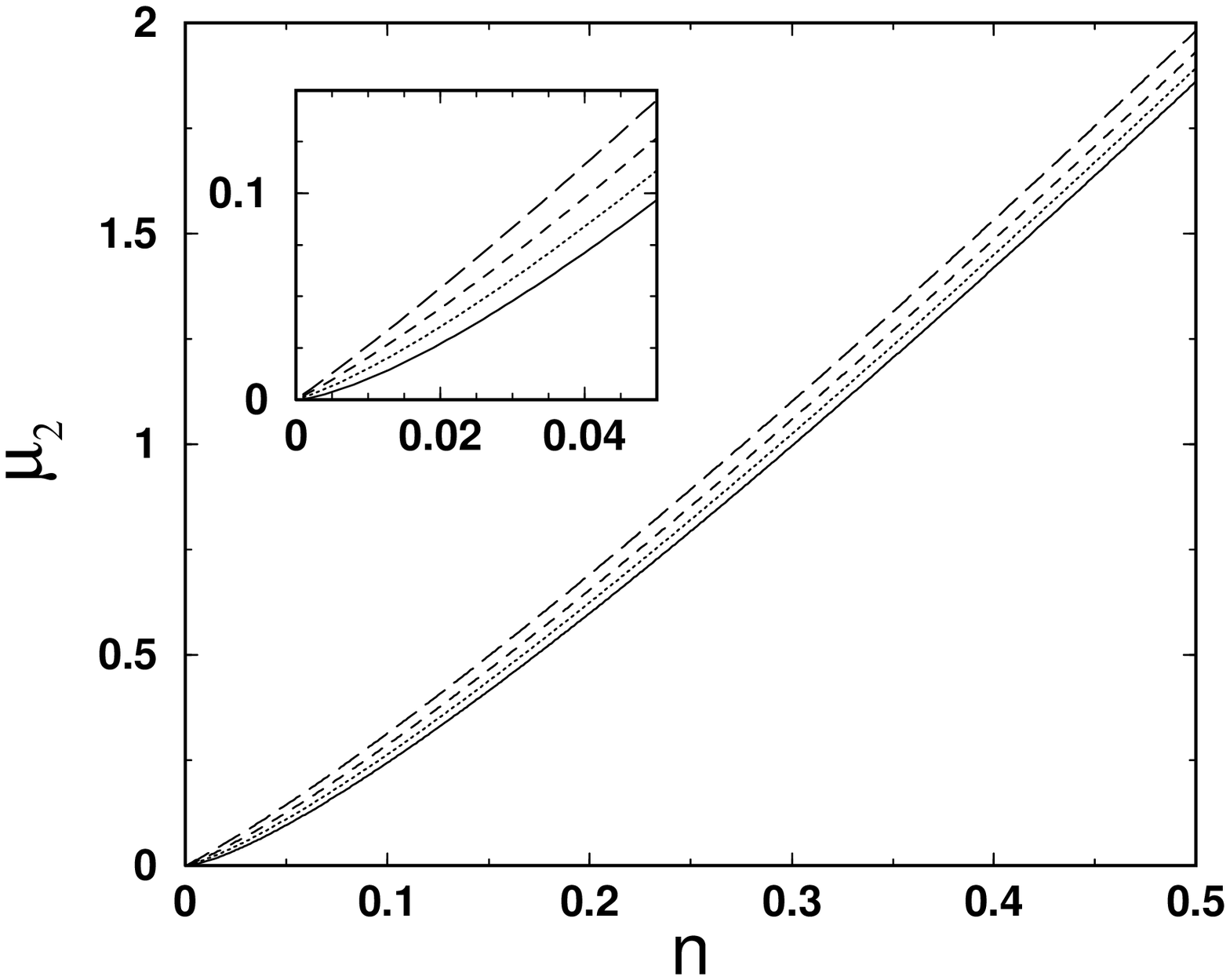}}
\caption{Second order contribution to the chemical potential $\mu_{2}$ 
vs density.
($\alpha=$ 0, 1, 1.5, and 2 -- solid, dotted, dashed and long dashed lines,
respectively,  $\omega_{0}=0.2$ and $g=3$.}
\label{f10}
\end{figure}
\begin{figure}
      \epsfysize=70mm
      \centerline{\epsffile{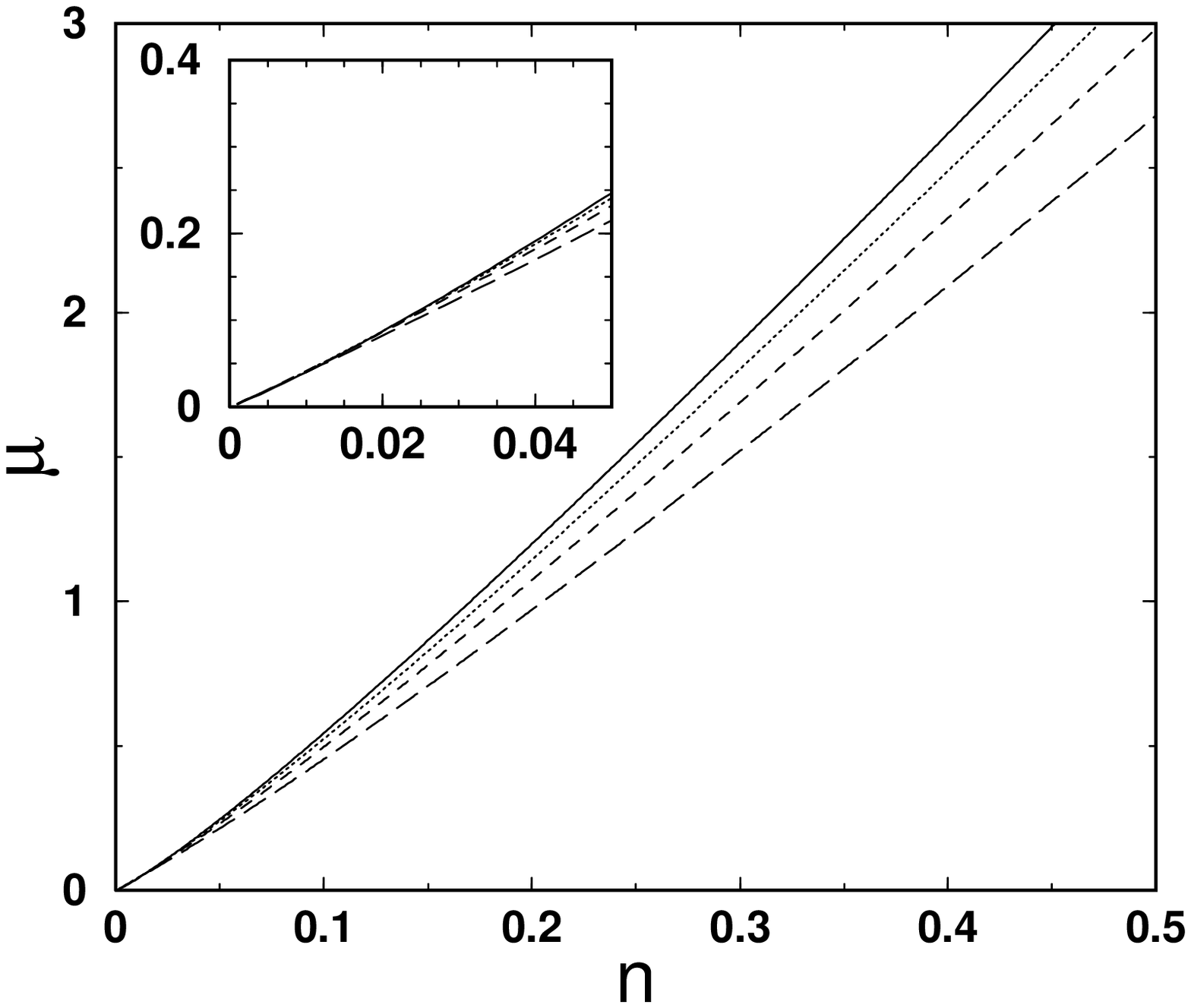}}
\caption{Chemical potential $\mu$ 
vs density $n$.
($\alpha=$0, 1, 1.5, and 2 -- solid, dotted, dashed and long dashed lines,
respectively,  $\omega_{0}=0.2$ and $g=3$).}
\label{f11}
\end{figure}
Keeping only  terms in leading order in the density, the expression for  
$\mu_{b}$ Eq.~(\ref{mubp}) becomes
\begin{eqnarray}
\mu_{b}&\simeq&\frac{g}{N}\sum_{\bf k}
\left\{\left[\frac{\bar{g}n_{c}+\varepsilon_{\bf k}-\bar{E}_{\bf k}}
{2\bar{E}_{\bf k}}+\frac{gn_{c}}{2}\left(\frac{1}{\varepsilon_{\bf k}}-
\frac{1}{\bar{E}_{\bf k}}\right)\right]\right.\nonumber\\
&+&\left.\alpha^2\omega_{0}^2n_{c}\frac{\omega_{0}+2\bar{E}_{\bf k}}{\bar{E}_{\bf k}(\omega_{0}+\varepsilon_{\bf k})^2}\right\}~.
\label{mub}
\end{eqnarray}
It is the contribution in square brackets in Eq.~(\ref{mub}),  which leads 
to the canonical $n^{3/2}$ behavior of the chemical potential. On the 
contrary, the second term, which is due to boson-phonon coupling, is 
linear in density. Upon increasing  $\alpha$ this latter contribution 
to the chemical potential increases 
while the term proportional to $n^{3/2}$  decreases, because
of the reduction of the effective scattering length $\bar{g}$.
This behavior is clearly seen from Fig. 11, where the chemical potential
$\mu=\mu_{1}+\mu_{2}$ is illustrated as a function of density.
For small $n$ the contribution linear in density is dominating and the
reduction of the  chemical potential due to the phonon-mediated effective
attraction in lowest order is almost canceled by the second order
contribution. As a result, the slope of $\mu$ is almost independent of 
$\alpha$ at small densities [see the inset in Fig. 11], while at higher values 
of $n$ the chemical potential decreases with increasing of $\alpha$, 
because of a decrease of the contribution proportional to $n^{3/2}$.

\subsection{The renormalized sound velocity}

The renormalized quasi-particle spectrum, arising from  the second order 
self-energy corrections, are given by the poles of the Boson Green's 
function (\ref{DB}),
or equivalently by the zeros of ${\text D}_{{\bf q},{\omega}}$ (\ref{DBnot}).
Expanding the self-energies in the long-wavelength and  low
frequency limit around the mean field spectrum, we obtain  the
renormalized sound velocity $v$ as
\begin{eqnarray}
v=v_{0}[1+\lambda].
\label{v}
\end{eqnarray}
$v_{0}=\sqrt{{\bar g}n_{c}/M}$ denotes the mean field sound velocity and
$\lambda$  the velocity renormalization factor given by
\begin{eqnarray}
\lambda=\hat{\Sigma}^{(2)}_{0,0}/(2\bar{g}n_{c})+\bar{g}n_{c}s-a.
\label{lambda}
\end{eqnarray}
$\hat{\Sigma}^{(2)}_{0,0}$ denotes the second order correction to the
anomalous self-energy. The symmetric and antisymmetric self-energies, 
to leading order in $\omega$ and ${\bf q}$ [see Eq.~(\ref{SE})], are
given by
\begin{eqnarray}
s=\lim_{{\bf q}\to 0}\frac{{\cal S}_{{\bf q},\omega_{1,{\bf q}}}-\mu}
{\omega_{1,{\bf q}}^{2}}, \; 
a=\lim_{{\bf q}\to 0}\frac{{\cal A}_{{\bf q},\omega_{1,{\bf q}}}}
{\omega_{1,{\bf q}}}.
\label{sa}
\end{eqnarray}

Each of the three contributions entering in sound velocity renormalization
factor, Eq.~(\ref{lambda}), contains infrared divergences, which are 
inherent to the Bose condensed state. However, as in the case of the 
standard weakly interacting Bose gas, these divergent terms are exactly 
canceled out in the final result for the low-frequency
excitation spectrum. We have checked this cancellation analytically, 
picking up the divergent terms both due to the coherence factors and the 
existence of the gapless mode. 

In Fig. 12 the sound velocity renormalization factor is presented for 
various values of the coupling constant $\alpha$. In the absence of 
boson-phonon coupling $\lambda$ follows the canonical  square root 
behavior in the density.\cite{SG} An increase of $\alpha$
results in an overall enhancement of $\lambda$ and moreover leads to a
finite value of $\lambda$,  even for vanishing density. As a result, 
the renormalized sound velocity [see Fig. 13] is almost unaffected by 
the coupling of the bosons with phonons, in contrast to its mean field 
result, which  is also presented in Fig. 13.
\begin{figure}
      \epsfysize=70mm
      \centerline{\epsffile{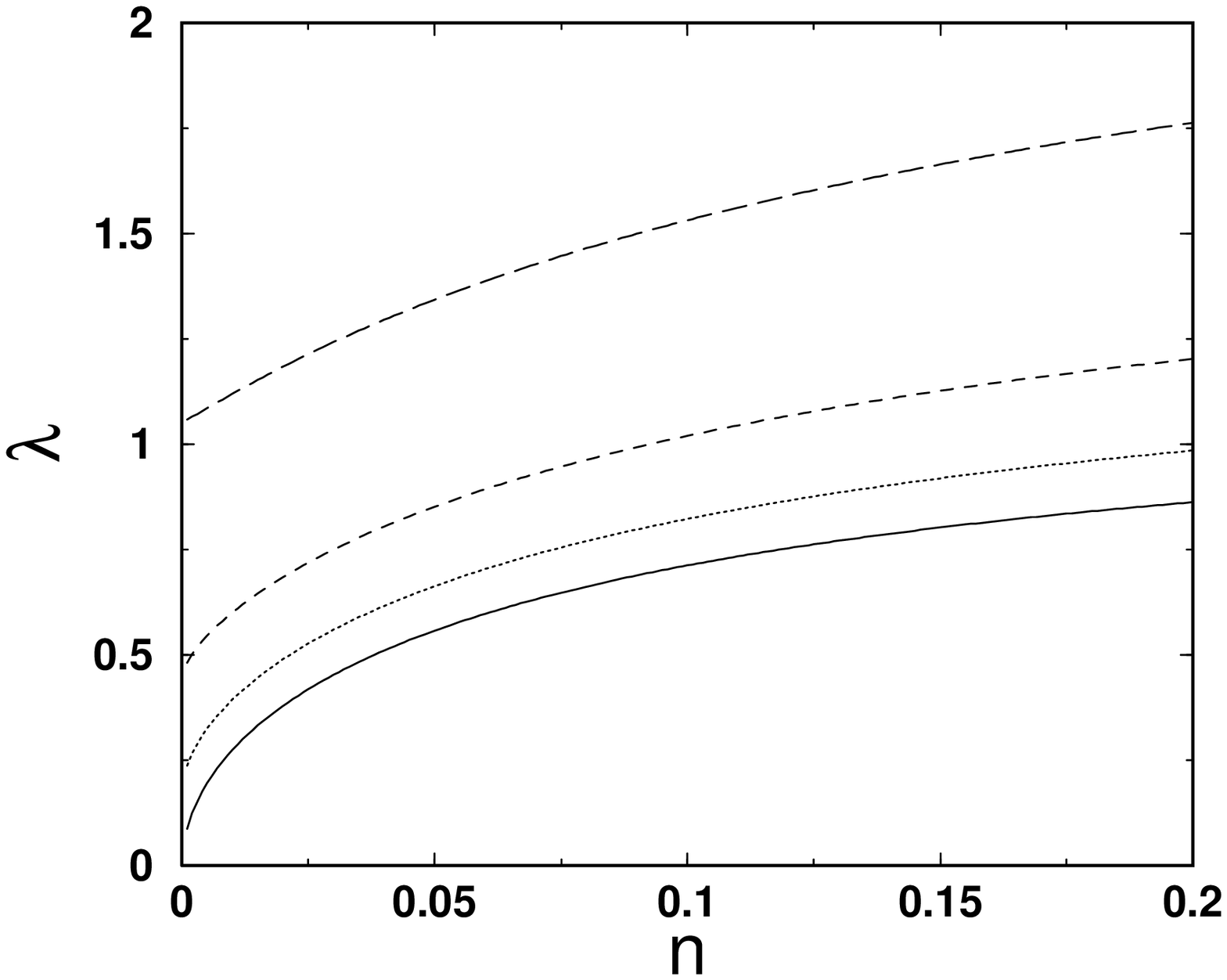}}
\caption{Velocity renormalization factor $\lambda$ 
versus density $n$.
($\alpha=$0,  1, 1.5, and 2 -- solid, dotted, dashed and long dashed lines,
respectively,  $\omega_{0}=0.2$ and $g=3$).}
\label{f12}
\end{figure}
\begin{figure}
      \epsfysize=70mm
      \centerline{\epsffile{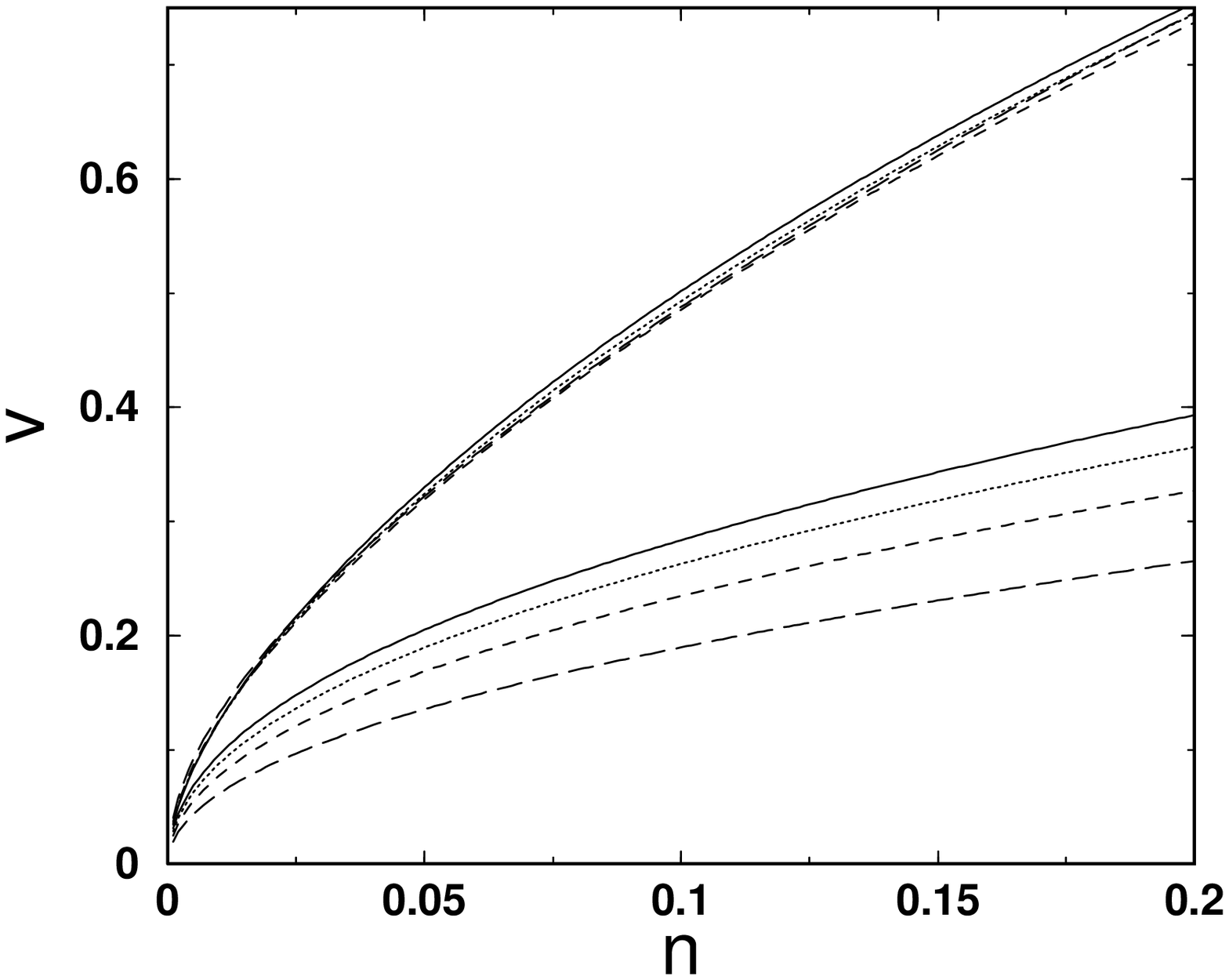}}
\caption{Renormalized sound velocity $v$
 (upper curves), mean-field sound velocity (lower curves) versus the 
density $n$ ($\alpha=$0, 1, 1.5, and 2 -- solid, dotted, dashed and 
long dashed lines,
respectively.  $\omega_{0}=0.2$ and $g=3$).}
\label{f13}
\end{figure}
The interesting question now is, why the boson dressing effects, 
leading to its mass
enhancement, does not show up in the velocity renormalization factor.
Responsible for the mass enhancement is the first diagram in Fig. 5 for
the normal self-energy. It is frequency dependent and is finite in the limit
$n\to 0$. It gives a positive contribution to $a$ and hence a negative 
contribution to the sound velocity [see Eq.~(\ref{lambda})]. It turns out 
that in the condensed state there are two effects that overcompensate this 
negative contribution. First, the effective 
boson-phonon coupling is no longer local and becomes momentum dependent
because of the coherence factors. It is decreased in the condensed state. 
For the momentum transfer equal to the wavevector 
at which the level crossing of the bare modes occurs, this effective 
interaction is exactly zero.  The second main important
contribution competing with this diagram is contained in the first diagram 
illustrated in Fig. 6 for
the anomalous self-energy. This diagram is positive, thus leading to a positive
contribution to the anomalous part of inter-particle potential, and hence 
increases the  sound velocity. 

In Fig. 14 we present these two contribution for various $\alpha$, rescaling
them by a factor $(\alpha^2\omega^2_0)^{-1}$. As can be  seen from this figure 
there is almost an exact cancellation of these two contributions for 
different $\alpha$. That explains the absence of the density independent 
negative contribution to the
sound velocity originating from the boson mass enhancement.
\begin{figure}
      \epsfysize=70mm
      \centerline{\epsffile{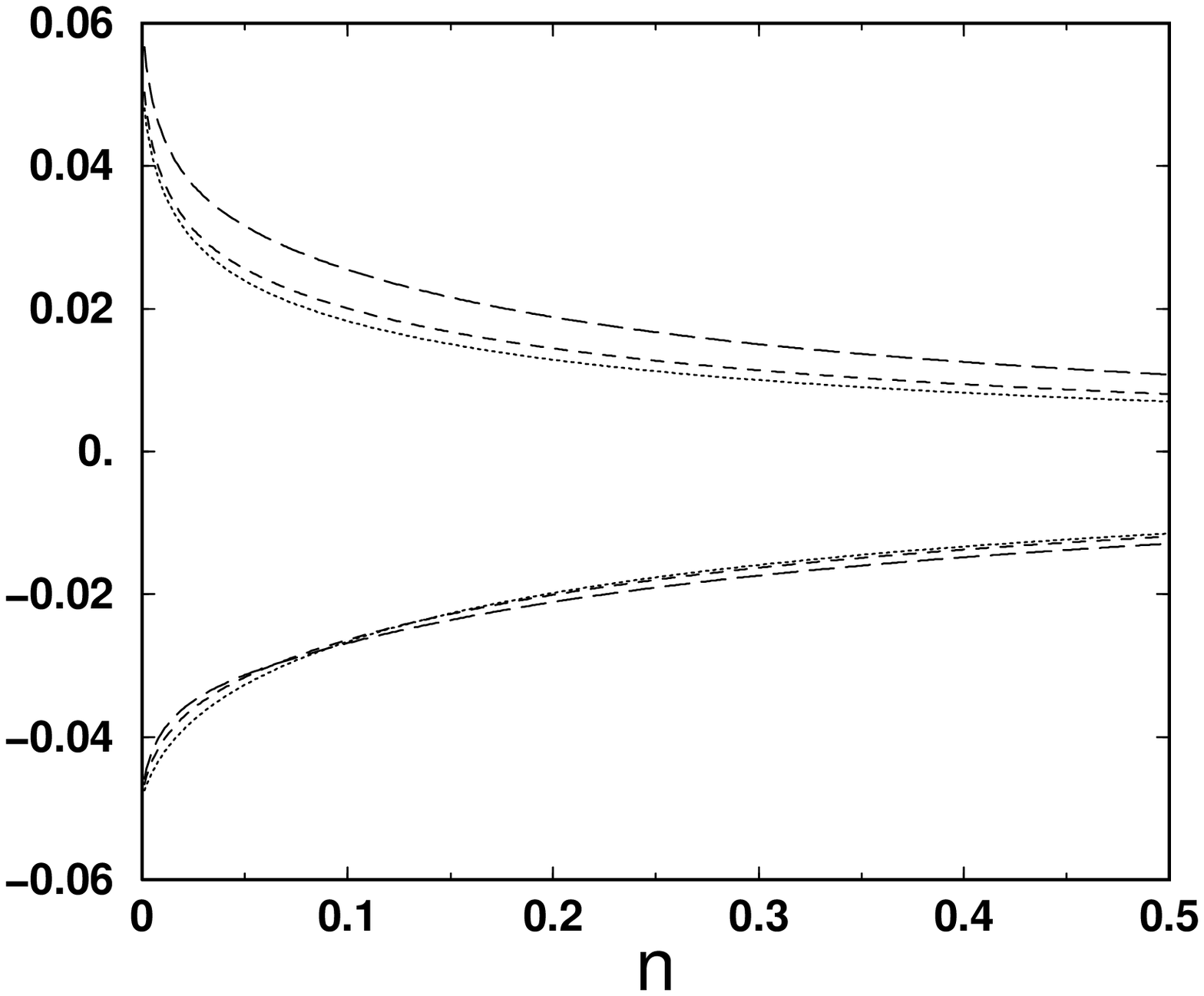}}
\caption{The negative (positive) contribution to the sound velocity 
renormalization factor from normal (anomalous) second order diagrams 
illustrated  by the first diagram in Fig.5 (Fig. 6)
($\alpha=$ 1, 1.5, and 2 -- dotted, dashed and long dashed lines,
respectively,  $\omega_{0}=0.2$ and $g=3$).}
\label{f14}
\end{figure}

\section{Conclusion}

In the present paper we have considered the problem of a system of bosons 
on a deformable lattice which, apart from an intrinsic repulsion between 
them, are coupled to local lattice deformations - treated in terms of Einstein 
phonon modes. This leads to a retarded time dependent contribution to the 
effective two-body vertex.  When the bosons condense, the phonons,  
being initially coupled to the boson density (symmetry restoring variable in 
the case of gauge symmetry breaking), get hybridized with the Goldstone mode.
The generalized Bogoliubov transformation has been used to diagonalize the 
bilinear part of Hamiltonian describing the excitation spectra of the system 
in the lowest order approximation. As a result two dispersion branches have 
been obtained: one describing a gapless sound-wave like mode and one exhibiting 
a gap. The two normal modes describe  Bogoliubov type  excitations
and Einstein phonons. For  momenta close to where the level crossing
of bare excitation spectra occurs neither mode is predominantly a phonon 
or a Bogoliubov quasi-particle. 

The ground state depletion is shown to consist of two contributions:
the first one  showing 
the canonical $n^{3/2}$ behavior in the density and the second one, 
which appears because of the retarded phonon mediated interaction,
being linear in density. The relative depletion of the ground state 
remains finite in the limit $n\to 0$ because of retardation effects. 

The sound velocity obtained in this approximation is reduced 
 due to reduction of the repulsive boson-boson interaction, arising from 
the attractive part of phonon mediated interaction in the static limit.

Considering the second order corrections to the chemical potential and 
excitation spectrum, the Beliaev-Popov theory has been generalized to our case.
For that purpose an  effective interaction has been introduced and all  second-order 
self-energy diagrams were constructed in terms of this interaction.
The short range bare boson-boson two-body potential has been renormalized in the
standard way by summing up the ladder diagrams and introducing the two-body
$t$-matrix with the characteristic $s$-wave scattering length. 
Since the phonon mediated boson-boson interaction, vanishes at high 
frequencies as $1/\omega^{2}$, this interaction does not require any 
special renormalization in order to avoid ultraviolet divergences in the case
of continuum model. Due to the pole structure of phonon mediated interaction,
the density of bosons no longer enters in the expansion parameter and one has 
to assume  a small value of boson-phonon coupling  constant
in order to treat the problem by perturbation theory.
  
Unlike the lowest order results, the second order contribution to 
both the chemical potential and the sound velocity shows an increase 
with the boson-phonon coupling constant and contains a term linear in the 
density which is  exclusively due to the retarded nature of phonon mediated 
interaction. 
As a result the total chemical potential and the sound velocity obtained within 
this theory are practically unaffected by the coupling to the phonons. 
This effect is more pronounced in the low density limit $n\ll 1$. 
One might ascribe this effect to the robustness 
of the superfluid state. 

In the present paper we have restricted ourself to the case
when the boson system is coupled to optical phonon modes of the lattice.
The formalism developed here
can be easily extended to the case of acoustic phonons. 
In such a scenario the two normal modes correspond to
 two types of sound modes: the acoustic phonon  
mode of the lattice and the Goldstone mode of condensed Bose system. 
The effect of the coupling between these two sound modes as well as
as well as damping effects can be treated within the present formalism.  
However,  in that case the coupling with $q=0$ phonon mode (representing 
translation of the whole crystal) should be discarded.
It amounts to neglecting the Hartree type contribution to 
the boson self-energy due to the phonon mediated interaction.

The present study induces us to speculate on the possibility of an 
insulator to superconducting transition when the boson-phonon coupling 
constant is strong. In that case the mass renormalization of the bosons 
varies exponentially with the coupling constant as $\exp(\alpha^2)$ when 
we consider the normal state of the system. In the superfluid state we expect 
again a phase stiffness of the condensate practically unaffected by the 
coupling of the bosons to the lattice vibrations. Possibly such features 
exist in the cuprate superconductors which show a resistivity which, upon 
lowering the temperature, tends to an insulating behavior before abruptly 
dropping to zero when the system becomes superconducting.
Similarly, upon entering the superconducting state, out of the broad 
incoherent contribution to the angle resolved photo-emission spectra  
evolves a sharp resonance peak which could suggest well defined quasi-particles 
in the superconducting state and totally diffusive modes in the normal state. 
Within the physics developed in the present study we can speculate on an 
undressing of the bosons as the temperature is lowered and the transition 
from the normal into the superfluid state takes place. These and related 
questions will be addressed in some future work.

\acknowledgments
One of the authors (G. J) acknowledges support from a  
{\it bourse de Recherche Scientifique et technique de l'OTAN} and from an 
 INTAS Program, Grant No 97-0963 and No 97-11066.
He acknowledges in particular the  kind hospitality at the
Centre de Recherches sur les Tr\`es Basses Temp\'eratures, where the main 
part of the present work has been carried out and the
Max-Plank Institute f$\ddot{\rm u}$r Physik komplexer Systeme, where the 
final part of the work has been completed. 

\end{multicols}
\widetext
\appendix
\section*{Effective vertices}
As we have already mentioned in the main text, each self-energy contribution
introduced in Eq.~(\ref{SE}) consist  of three different parts
describing the intra-band (denoted as 11(22) for the scattering within
the lower (upper) branch) and inter-band (denoted as 12) scattering processes.
The vertices for inter-band scattering are given by

\begin{eqnarray}
A_{12}({\bf k},{\bf q})&=&4g^{2}n_{c}\Bigl[
(u_{1,{\bf k}}^2{\bar u}_{1,{\bf k+q}}^2
-v_{1,{\bf k}}^2{\bar v}_{1,{\bf k+q}}^2)-2u_{1,{\bf k}}v_{1,{\bf k}}
({\bar u}_{1,{\bf k+q}}^2-{\bar v}_{1,{\bf k+q}}^2)-2{\bar u}_{1,{\bf
k+q}}{\bar v}_{1,{\bf k+q}}(u_{1,{\bf k}}^2-v_{1,{\bf k}}^2)\Bigr]
\nonumber\\
&-&4g\alpha\omega_{0}\sqrt{n_c}\Bigl[(u_{2,{\bf k}}-v_{2,{\bf k}})
\bigl\{(u_{1,{\bf k}}-v_{1,{\bf k}})({\bar u}_{1,{\bf k+q}}^2-
{\bar v}_{1,{\bf k+q}}^2)
-(u_{1,{\bf k}}+v_{1,{\bf k}}){\bar u}_{1,{\bf k+q}}{\bar v}_{1,{\bf k+q}}
\bigr\}
\Bigr.\nonumber\\
&+&({\bar u}_{2,{\bf k+q}}-{\bar v}_{2,{\bf k+q}})
\Bigl.\bigl\{({\bar u}_{1,{\bf
k+q}}-{\bar v}_{1,{\bf k+q}})(u_{1,{\bf k}}^2-v_{1,{\bf k}}^2)-({\bar
u}_{1,{\bf k+q}}+{\bar v}_{1,{\bf k+q}})u_{1,{\bf k}}v_{1,{\bf
k}}\bigr\}\Bigr]\nonumber\\
&+&8g\alpha^2\omega_{0}n_{c}\Bigl[u_{1,{\bf k}}v_{1,{\bf k}}
({\bar u}_{1,{\bf k+q}}^2-{\bar v}_{1,{\bf k+q}}^2)+
{\bar u}_{1,{\bf k+q}}{\bar v}_{1,{\bf k+q}}
(u_{1,{\bf k}}^2-v_{1,{\bf k}}^2)\Bigr]
\nonumber\\
&-&2\alpha^3\omega_{0}^2\sqrt{n_{c}}\Bigl[\bigl\{(u_{2,{\bf k}}-
v_{2,{\bf k}})({\bar
u}_{1,{\bf k+q}}+{\bar v}_{1,{\bf k+q}})+({\bar
u}_{2,{\bf k+q}}-{\bar v}_{2,{\bf k+q}})(u_{1,{\bf k}}+v_{1,{\bf
k}})\bigr\}\bigl\{
u_{1,{\bf k}}{\bar v}_{1,{\bf k+q}}+{\bar u}_{1,{\bf k+q}} v_{1,{\bf
k}} \bigr\}\Bigr]\nonumber\\
&+&\alpha^2\omega_{0}^{2}\Bigl[(u_{2,{\bf k}}-v_{2,{\bf k}})^2({\bar u}_{1,{\bf
k+q}}^2-{\bar v}_{1,{\bf k+q}}^2)+({\bar u}_{2,{\bf k+q}}-
{\bar v}_{2,{\bf k+q}})^2(u_{1,{\bf k}}^2-v_{1,{\bf k}}^2)
\Bigr.\nonumber\\
&+&\Bigl.
2(u_{2,{\bf k}}-v_{2,{\bf k}})({\bar u}_{2,{\bf
k+q}}-{\bar v}_{2,{\bf k+q}})(u_{1,{\bf k}}{\bar u}_{1,{\bf
k+q}}-v_{1,{\bf k}}{\bar v}_{1,{\bf k+q}})\Bigr]\nonumber\\
S_{12}({\bf k},{\bf q})&=&4g^{2}n_{c}\Bigl[u_{1,{\bf k}}
{\bar u}_{1,{\bf k+q}}-v_{1,{\bf k}}{\bar v}_{1,{\bf
k+q}}\Bigr]^2 \nonumber\\
&-&4g\alpha\omega_{0}\sqrt{n_c}
\Bigl[(u_{1,{\bf k}}{\bar u}_{1,{\bf k+q}}-v_{1,{\bf k}}
{\bar v}_{1,{\bf k+q}})
\bigl\{(u_{2,{\bf k}}-v_{2,{\bf k}})({\bar u}_{1,{\bf k+q}}+
{\bar v}_{1,{\bf k+q}})+({\bar u}_{2,{\bf k+q}}-
{\bar v}_{2,{\bf k+q}})(u_{1,{\bf k}}+v_{1,{\bf k}})\bigr\}
\Bigr]\nonumber\\
&+&\alpha^2\omega_{0}^{2}\Bigl[(u_{2,{\bf k}}-v_{2,{\bf k}})^2
({\bar u}_{1,{\bf k+q}}+{\bar v}_{1,{\bf k+q}})^2+
({\bar u}_{2,{\bf k+q}}-{\bar v}_{2,{\bf k+q}})^2({ u}_{1,{\bf k}}
+{v}_{1,{\bf k}})^2\Bigr]\nonumber\\
M_{12}({\bf k},{\bf q})&=&8g^{2}n_{c}\Bigl[(u_{1,{\bf k}}^2
{\bar v}_{1,{\bf k+q}}^2
+{\bar u}_{1,{\bf k+q}}^2v_{1,{\bf k}}^2)-u_{1,{\bf k}}v_{1,{\bf k}}
({\bar u}_{1,{\bf k+q}}^2+{\bar v}_{1,{\bf k+q}}^2)\Bigr.\nonumber\\
&-&\Bigl.2{\bar u}_{1,{\bf
k+q}}{\bar v}_{1,{\bf k+q}}(u_{1,{\bf k}}^2+v_{1,{\bf k}}^2)
+3u_{1,{\bf k}}v_{1,{\bf k}}{\bar u}_{1,{\bf k+q}}{\bar v}_{1,{\bf
k+q}}\Bigr]
\nonumber\\
&+&4g\alpha\omega_{0}\sqrt{n_c}\Bigl[2\bigl\{(u_{1,{\bf k}}-
v_{1,{\bf k}})(u_{2,{\bf k}}-v_{2,{\bf k}}){\bar u}_{1,{\bf k+q}}
{\bar v}_{1,{\bf k+q}}+({\bar u}_{1,{\bf
k+q}}-{\bar v}_{1,{\bf k+q}})({\bar u}_{2,{\bf k+q}}-{\bar v}_{2,{\bf
k+q}})u_{1,{\bf k}}v_{1,{\bf k}}\bigr\}\Bigr.\nonumber\\
&+&\Bigl.\bigl\{(u_{2,{\bf k}}-v_{2,{\bf k}})({\bar u}_{1,{\bf
k+q}}-{\bar v}_{1,{\bf k+q}})+({\bar u}_{2,{\bf k+q}}-{\bar v}_{2,{\bf
k+q}})(u_{1,{\bf k}}-v_{1,{\bf k}})\bigr\}(u_{1,{\bf k}}{\bar v}_{1,{\bf k+q}}
+{\bar
u}_{1,{\bf k+q}}{ v}_{1,{\bf k}})\Bigr]\nonumber\\
&-&8g\alpha^2\omega_{0}n_{c}\Bigl[2(u_{1,{\bf k}}{\bar v}_{1,{\bf
k+q}}+{\bar u}_{1,{\bf k+q}}v_{1,{\bf k}})^2-u_{1,{\bf k}}v_{1,{\bf k}}
({\bar u}_{1,{\bf k+q}}^2+{\bar v}_{1,{\bf k+q}}^2)-
{\bar u}_{1,{\bf k+q}}{\bar v}_{1,{\bf k+q}}
(u_{1,{\bf k}}^2+v_{1,{\bf k}}^2)\Bigr]\nonumber\\
&-&2\alpha^3\omega_{0}^2\sqrt{n_{c}}\Bigl[(u_{1,{\bf k}}-
v_{1,{\bf k}})(u_{2,{\bf k}}-v_{2,{\bf k}}){\bar u}_{1,{\bf k+q}}
{\bar v}_{1,{\bf k+q}}+({\bar u}_{1,{\bf k+q}}-{\bar v}_{1,{\bf k+q}})
({\bar u}_{2,{\bf k+q}}-{\bar v}_{2,{\bf k+q}})u_{1,{\bf k}}
v_{1,{\bf k}}\Bigr.\nonumber\\
&+&\Bigl.(u_{2,{\bf k}}-v_{2,{\bf
k}})({\bar u}_{1,{\bf k+q}}^2v_{1,{\bf k}}-{\bar v}_{1,{\bf k+q}}^2
u_{1,{\bf k}})+({\bar u}_{2,{\bf k+q}}-{\bar v}_{2,{\bf k+q}})
(u_{1,{\bf k}}^2{\bar v}_{1,{\bf k+q}}-v_{1,{\bf k}}^2
{\bar u}_{1,{\bf k+q}})\Bigr]\nonumber\\
&+&8\alpha^4\omega_{0}^2n_{c}\Bigl[u_{1,{\bf k}}{\bar v}_{1,{\bf k+q}}+
{\bar u}_{1,{\bf k+q}}v_{1,{\bf k}}\Bigr]^2\nonumber\\
&-&2\alpha^2\omega_{0}^2
\Bigl[(u_{2,{\bf k}}-v_{2,{\bf k}})^2{\bar u}_{1,{\bf k+q}}
{\bar v}_{1,{\bf k+q}}+({\bar u}_{2,{\bf k+q}}-{\bar v}_{2,{\bf k+q}})^2
u_{1,{\bf k}}v_{1,{\bf k}}\Bigr.\nonumber\\
&+&\Bigl.(u_{2,{\bf k}}-v_{2,{\bf k}})({\bar u}_{2,{\bf k+q}}-{\bar
v}_{2,{\bf k+q}})(u_{1,{\bf k}}{\bar v}_{1,{\bf k+q}}+v_{1,{\bf k}}
{\bar u}_{1,{\bf k+q}})\Bigr]~.
\label{vert}
\end{eqnarray}
The corresponding vertices for the transitions within the lower 
(upper) band are obtained from the above expressions by replacing 
${\bar u},{\bar v}\rightarrow u,v$ ($u,v\rightarrow{\bar u},{\bar v}$) 
and multiplying by a factor $1/2$.

\begin{multicols}{2}

\end{multicols}

\begin{references}
\bibitem[\dagger]{byline} On leave of absence from Institute of Physics,
Georgian Academy of Sciences,  380077 Tbilisi, Georgia.
\bibitem{GRB} A. Griffin, {\it Excitations in a Bose-Condensed Liquid}
(Cambridge University Press, 1993).
\bibitem{EXEXP} A. Mysyrowicz, E. Benson, and E. Fortin,
Phys. Rev. Lett. {\bf 77}, 896 (1996);
T. Goto, M. Y. Shen, S. Koyama, and T. Yokuchi,
Phys. Rev. B {\bf 55}, 7609 (1997). 
\bibitem{Loutsenko-97} I. Loutsenko and D. Roubtsov, Phys. Rev. Lett. 
{\bf 78}, 3011 (1997).
\bibitem{Ionexp} R.P. Sharma, T. Venkatesan, Z.H. Zhang, J. R. Liu, R Chu 
and W. K. Chu, Phys. Rev. Lett. {\bf 77}, 4624 (1996).
\bibitem{Raman} G. Ruani and P. Ricci, Phys. Rev. B {\bf 55}, 93 (1997); 
O. V. Misochko, E. Ya. Sherman, N. Umesaki and S. Nakashima, Phys. Rev. B 
{\bf 59}, 11495 (1999).
\bibitem{Bogoliubov} N. N. Bogoliubov, J. Phys. USSR {\bf 5}, 23 (1947).
\bibitem{B} S. T. Beliaev, Sovjet Physics JETP {\bf 7}, 299 (1958). 
\bibitem{P}
V.N. Popov,
``Functional Integrals in Quantum Field theory and Statistical Physics'', 
Ch. 6, Reidle, Dortrecht (1983)
\bibitem{GN} J. Gavoret and P. Nozi\`eres, Ann. Phys. (New York) {\bf 28},
349 (1964).
\bibitem{SG} H. Shi and A. Griffin, Phys. Rep. {\bf 304}, 1 (1998).
\bibitem{HP} N. M. Hugenholtz and D. Pines, Phys. Rev. {\bf 116}, 489 (1959).
\bibitem{GRIF} A. Griffin, Phys. Rev. {\bf B 53}, 9341 (1996). 
\end{references}
\end{document}